\documentclass[journal=jctcce,manuscript=article]{achemso}

\usepackage[version=3]{mhchem} 
\usepackage{amsmath,amssymb,amsthm, amsfonts, mathtools, dsfont} 
\usepackage{bbm,bm,tensor, braket}
\usepackage{eqnarray,array,enumerate}
\usepackage{csquotes}
\usepackage{siunitx}
\usepackage{booktabs}
\usepackage{graphicx,wrapfig, caption, float, subcaption, epstopdf, setspace}
\usepackage{hyperref}
\usepackage{longtable}
\usepackage[section]{placeins}
\usepackage{multicol, multirow}
\usepackage{pgfplots}
\pgfplotsset{compat=1.9}
\usepgfplotslibrary{fillbetween}
\usepackage{algorithm, algpseudocode}

\usepackage{cleveref}

\newcommand*{\citen}[1]{%
  \begingroup
    \romannumeral-`\x 
    \setcitestyle{numbers}%
    \cite{#1}%
  \endgroup   
}
\usepackage{mathabx}
\def\br{\ensuremath\bm{r}}
\author{Arno Förster}
\email{a.t.l.foerster@vu.nl}
\affiliation{Theoretical Chemistry, Vrije Universiteit, De Boelelaan 1083, NL-1081 HV, Amsterdam, The Netherlands}

\author{Erik van Lenthe}
\email{vanlenthe@scm.com}
\affiliation{Software for Chemistry and Materials NV, NL, 1081HV, Amsterdam, The Netherlands}

\author{Edoardo Spadetto}
\affiliation{Software for Chemistry and Materials NV, NL, 1081HV, Amsterdam, The Netherlands}

\author{Lucas Visscher}
\affiliation{Theoretical Chemistry, Vrije Universiteit, De Boelelaan 1083, NL-1081 HV, Amsterdam, The Netherlands}

\title{Two-component $GW$ calculations: Cubic scaling implementation and comparison of vertex corrected and partially self-consistent $GW$ variants}


\DeclareUnicodeCharacter{2212}{-}
\begin{document}

\begin{abstract}
We report an all-electron, atomic orbital (AO) based, two-component (2C) implementation of the $GW$ approximation (GWA) for closed-shell molecules. Our algorithm is based on the space-time formulation of the GWA and uses analytical continuation of the self-energy, and pair-atomic density fitting (PADF) to switch between AO and auxiliary basis. By calculating the dynamical contribution to the $GW$ self-energy at a quasi-one-component level, our 2C $GW$ algorithm is only about a factor of two to three slower than in the scalar relativistic case. Additionally, we present a 2C implementation of the simplest vertex correction to the self-energy, the statically screened $G3W2$ correction. Comparison of first ionization potentials of a set of 67 molecules with heavy elements (a subset of the SOC81 set) calculated with our implementation against results from the WEST code reveals mean absolute deviations of around 70 meV for $G_0W_0$@PBE and $G_0W_0$@PBE0. These are most likely due to technical differences in both implementations, most notably the use of different basis sets, pseudopotential approximations, different treatment of the frequency dependency of the self-energy and the choice of the 2C-Hamiltonian. However, how much each of these differences contribute to the observed discrepancies is unclear at the moment. Finally, we assess the performance of some (partially self-consistent) variants of the GWA for the calculation of first IPs by comparison to vertical experimental reference values. $G_0W_0$PBE0 (25 \% exact exchange) and $G_0W_0$BHLYP (50 \% exact exchange) perform best with mean absolute deviations (MAD) of about 200 meV. Explicit treatment of spin-orbit effects at the 2C level is crucial for systematic agreement with experiment. On the other hand eigenvalue-only self-consistent $GW$ (ev$GW$) and quasi-particle self-consistent $GW$ (qs$GW$) significantly overestimate the IPs. Perturbative $G3W2$ corrections increase the IPs and therefore improve the agreement with experiment in cases where $G_0W_0$ alone underestimates the IPs. With a MAD of only 140 meV, 2C-$G_0W_0$PBE0 + $G3W2$ is in best agreement with the experimental reference values. 
\end{abstract}

\maketitle

\section{Introduction}
Due to its favorable price-to-performance ratio, the $GW$ approximation (GWA)\cite{Hedin1965, martin2016} ($G$: single-particle Green's function, $W$: screened electron-electron interaction) is one of the most popular methods for the calculation of charged excitations in finite systems.\cite{Reining2018,Golze2019} Over the last decade, the GWA has been implemented into a large number of electronic structure codes\cite{Ren2012, Caruso2012, Caruso2013,
VanSetten2013,
Kaplan2015,
Kaplan2016,
Bruneval2016a,
Foerster2011a,Koval2014,
Mejia-Rodriguez2021,Forster2020b,Forster2021a,
Wilhelm2016a, Wilhelm2018, Wilhelm2021,
Ke2011} and $GW$ implementations for massively parallel architectures,\cite{Govoni2015, Wilhelm2016a, DelBen2019, DelBen2019a, Yu2022} low-order scaling implementations,\cite{Wilhelm2018, Wilhelm2021, Forster2020b, Forster2021a, Duchemin2021a} effectively linear scaling stochastic formulations,\cite{Vlcek2017, Vlcek2018} fragment-based approaches\cite{Fujita2018, Fujita2019, Winter2021, Amblard2022} or embedding techniques\cite{Romanova2020, Weng2021,Tolle2021} have enabled applications of the $GW$ method to large biomolecules,\cite{Forster2021a,Forster2022c} nanostructures\cite{BorinBarin2022, Amblard2022, Yu2022} or interfaces.\cite{Yu2022} 

A large numbers of studies has by now contributed to a thorough understanding of the impact of technical aspects of these implementations, like the choice of single-particle basis, pseudopotential (PP) approximations, or frequency treatment,\cite{VanSetten2015, Maggio2016, Govoni2018, Gao2019,Bruneval2020,Forster2021a} as well as the performance of various $GW$ approaches for the first ionization potentials (IP) and electron affinities (EA) of weakly correlated organic molecules.\cite{Bruneval2009, Marom2012, Bruneval2013, Ren2015, Knight2016, Rangel2016, Caruso2016, Forster2022} More recently, the GWA has also been benchmarked for core excitations\cite{Golze2018, VanSetten2018a, Golze2020, Yao2022, Li2022a} and strongly correlated systems like open-shell molecules\cite{Mansouri2021} or transition metal compounds with partially filled 3$d$ shells.\cite{Korbel2014, Berardo2017, Hung2017, Shi2018, Byun2019, Rezaei2021, Wang2022} Fully self-consistent $GW$ (sc$GW$) calculations are relatively expensive, technically demanding, and not necessarily very accurate for the calculation of IPs and EAs.\cite{Marom2012, Caruso2016, Knight2016} Instead, the much cheaper perturbative $G_0W_0$ approach\cite{Hybertsen1985, Hybertsen1986} or its eigenvalue-only self-consistent extension (ev$GW$) are typically the method of choice. Despite their often excellent accuracy, these methods fail when the KS orbitals for which the $GW$ corrections are evaluated are qualitatively wrong.\cite{Bruneval2013, Knight2016, Forster2022c} In the quasi-particle self-consistent $GW$ method (qs$GW$),\cite{Faleev2004, VanSchilfgaarde2006, Kotani2007} the frequency dependent and non-Hermitian $GW$ self-energy is mapped self-consistently to an effective static and Hermitian non-local potential which is a functional of the non-interacting single-particle Green's function. Therefore, the results are strictly independent of the KS density functional which is used as starting-point for the calculation.\cite{Forster2021a, Forster2022c} The available benchmark data suggest that for molecules qs$GW$ is at least as accurate as $G_0W_0$.\cite{Ke2011, Gui2018, Forster2022} 

Less is known about the accuracy of the GWA for molecules containing heavier elements. One reason for this is that for those systems only a limited number of accurate first-principle results are available.\cite{Akinaga2017, Shee2018} Another reason is that comparison to experimental data is complicated by spin-orbit coupling (SOC) whose explicit treatment requires to implement the GWA in a 2-component (2C) framework. While Aryasetiawan and coworkers have generalized Hedin's equation to spin-dependent interactions\cite{Aryasetiawan2008, Aryasetiawan2009} more than a decade ago, only a few 2C implementations of the GWA for molecules have been realized so far.\cite{Kuhn2015, Scherpelz2016, Holzer2019, Franzke2022, Holzer2023} The probably most systematic study of SOC effects in molecules has been performed by Scherpelz and Govoni\cite{Scherpelz2016} who have compiled a set of 81 molecules containing heavy elements (referred to as SOC81 in the following).\cite{Scherpelz2016} They performed two-component (2C) $GW$@PBE\cite{Perdew1996} and $GW$@PBE0\cite{Adamo1999, Ernzerhof1999} calculations for this set using the WEST code\cite{Govoni2015, Yu2022} and found that SOC can shift scalar relativistic (1C) first ionization potentials by up to 400 meV for molecules containing iodine.\cite{Scherpelz2016} Interestingly, they observed that the 1C results were often closer to experiment than the 2C ones. Also, the fact that $GW$@PBE and $GW$@PBE0 are not necessarily very accurate for molecules\cite{Caruso2016,Knight2016,Wang2021,Zhang2022} suggests that the good performance of those methods for these systems might at least partially be due to fortuitous error cancellation. The accuracy of $G_0W_0$ calculations based on starting points with a higher fraction of exact exchange has however not been systematically investigated for molecules containing heavy elements. Also, little is known about the performance of partially self-consistent approaches.

In efforts to improve over the $GW$ approximation, also the role of higher order terms in the expansion of the electronic self-energy in terms of $W$ (vertex corrections), has been assessed over the last years for small and medium molecules.\cite{Ren2015, Ma2019a, Vlcek2019, Pavlyukh2020, Bruneval2021a, Wang2021, Wang2022a, Forster2022, Mejuto-Zaera2022a} The available results suggest that they generally fail to improve consistently over the best available $GW$ variants when they are combined with QP approximations.\cite{Kutepov2017, Kutepov2017a, Forster2022} However, they can remove some of the starting point dependence of $G_0W_0$\cite{Wang2021, Wang2022a} and often tremendously improve the description of electron affinities.\cite{Vlcek2019, Forster2022a} With the exception of one recent study which focused on first-row transition metal oxides,\cite{Wang2022a} the available benchmark results are limited to charged valence excitations in mostly organic molecules. It is not known how these methods perform for molecules containing heavier elements, where electron correlation effects and screening effects might be stronger.

In this work, we address some of these open questions. We present systematic benchmarks of 2C-GWA at different levels of self-consistency, ranging from $G_0W_0$ to qs$GW$. We also investigate the effect of the statically screened $G3W2$ term\cite{Forster2022} on the QP energies in a 2C framework. Our calculations are performed using a newly developed 2C (qs)$GW$ implementation, a generalization of our atomic orbital based qs$GW$ and $G_0W_0$ algorithms.\cite{Forster2020b,Forster2021a} Our 2C implementation retains the same favorable scaling with system size and increases the prefactor of the calculations by only a factor of two compared to the 1C case. This relatively small increase in computational effort is achieved by calculating the dynamical contributions to the electron self-energy at a  quasi-one-component level. Therefore, our new implementation also allows us to describe SOC effects in large molecules. All other quantities, including the polarizability, are treated at the full 2C level without any further approximations.

The remainder of this paper is organized as follows: In section~\ref{sec::theory}, we review the 2C-$GW$ working equations and give a detailed overview of our implementation. After describing the details of our calculations in section~\ref{sec::computationalDetails}, we report the results of our detailed benchmark calculations in section~\ref{sec::theory}: First, to assess the influence of the different technical parameters in both implementations, we compare $G_0W_0$@PBE0 IPs for SOC81 to the ones from Scherpelz and Govoni\cite{Scherpelz2016}. We then use our new implementation to calculate the first ionization potentials of the molecules in the SOC81 database using some of the most accurate available $GW$ approaches: qs$GW$, eigenvalue-only self-consistent $GW$ (ev$GW$), eigenvalue-only self-consistent $GW$ with fixed screened interaction after the first iteration (ev$GW_0$), and $G_0W_0$ based on hybrid starting points with different fractions of exact exchange. Finally, section~\ref{sec::conclusions} summarizes and concludes this work.

\section{\label{sec::theory}Theory}
\subsection{$GW$ approximation and $G3W2$ correction}
The central object of this work is the $GW + G3W2$ self-energy,
\begin{equation}
\label{self-energy-full}
    \Sigma^{GW + G3W2}(1,2) = 
    \Sigma_H(1,2) + \Sigma^{GW}(1,2) + 
    \Sigma^{G3W2}(1,2) \;.
\end{equation}
Here, 
\begin{equation}
    \Sigma_H(1,2) = v_H(1)\delta(1,2)
    = -i \delta(1,2)\int d3\; v_c(1,3)G(3,3^+) \;,
\end{equation}
with the Hartree-potential $v_H$,
\begin{equation}
\label{gw-self-energy-full}
    \Sigma^{GW}(1,2) = i G(1,2)W(1,2)  
\end{equation}
and    
\begin{equation}
\label{g3w2-self-energy-full}
\Sigma^{G3W2}(1,2) = -
\int d3 d4 G(1,3)W(1,4)G(3,4)G(4,2) W(3,2) \;.
\end{equation}
Space, spin, and imaginary time indices are collected as $1 = (\bm{r}_1,\sigma_1,i\tau_1)$. $W$ is the screened Coulomb interaction which is obtained by the Dyson equation \begin{equation}
\label{screened-coulomb-2}
   W(1,2) = W^{(0)}(1,2) + \int d3 d4 W^{(0)}(1,3)
   P^{(0)}(3,4)W(4,2) \;.
\end{equation}
Here, 
\begin{equation}
    W^{(0)}(1,2) = v_c(\bm{r}_1,\bm{r}_2)\delta_{\sigma,\sigma'}\delta(t_1-t_2) \;,
\end{equation}
is the bare Coulomb interaction and $P^{(0)}$ is the polarizability in the random phase approximation (RPA),
\begin{equation}
\label{chi0}
    P^{(0)}(1,2) = -i G(1,2)G(2,1) \;.
\end{equation}
Finally, $G$ is the interacting single-particle Green's function which is connected to its non-interacting counterpart $G^{(0)}$ by a Dyson equation with the electronic self-energy \eqref{self-energy-full} as its kernel,
\begin{equation}
\label{dyson}
    G(1,2) = G^{(0)}(1,2) + \int d3 d4 G^{(0)}(1,3) \Sigma(3,4) G(4,2) \;.
\end{equation}
If necessary, one can transform all quantities to imaginary frequency using the Laplace transform\cite{Rieger1999}
\begin{equation}
\label{TtoW}
    f(i\omega) = -i \int d\tau F(i\tau) e^{i\omega \tau} \;.
\end{equation}
The self-consistent solution of \cref{gw-self-energy-full,screened-coulomb-2,chi0,dyson} is referred to as $GW$ approximation. 

Typically, \eqref{dyson} is approximated. To this end, one defines an auxiliary Green's function $G^{(s)}$ which is related to $G^{(0)}$ by  
\begin{equation}
\label{dyson-aux}
    G^{(s)} = G^{(0)}(1,2) + \int d3 d4 G^{(0)}(1,3) v_{Hxc}(3,4) G^{(s)}(4,2) \;,
\end{equation}
where $v_{Hxc}$ is a (potentially local) generalized Kohn-Sham\cite{Hohenberg1964,Kohn1965, Seidl1996} Hartree-exchange-correlation potential. $G$ is then obtained from $G^{(s)}$ by 
\begin{equation}
\label{dyson-approx}
    G(1,2) = G^{(s)}(1,2) + \int d3 d4 G^{(s)}(1,3) \left[\Sigma_{Hxc}(3,4) - v_{Hxc}(3,4)\right]G(4,2) \;.
\end{equation}
In the basis of molecular orbitals (MO) $\left\{\phi_k\right\}$, $G^{(s)}$ is diagonal, 
\begin{equation}
\label{g_time-ordered_mo}
G^{(s)}_{pp'} = \Theta(i\tau)G^{>}_{pp'}(i\tau) - \Theta(-i\tau)G^{<}_{pp'}(i\tau) \;,
\end{equation}
with greater and lesser propagators being defined as 
\begin{equation}
\label{g_g_mo}
G^{>}_{pp'}(i\tau) = -i \Theta(\epsilon_p) e^{-\epsilon_p\tau}
\end{equation}
and
\begin{equation}
\label{g_l_mo}
G^{<}_{pp'}(i\tau) = -i \Theta(-\epsilon_p) e^{-\epsilon_p\tau} \;.
\end{equation}
Here, it is understood that all QP energies $\epsilon_k$ and KS eigenvalues $\epsilon^{KS}_k$ are measured relative to the chemical potential $\mu$ which we place in the middle of the HOMO-LUMO gap. $\Theta$ is the Heaviside step-function and $p,q,r,s \dots$ denote spinors. Under the assumption that the KS eigenstates are a good approximation to the $GW$ eigenstates, the off-diagonal elements of the operator $\Sigma_{Hxc} - v_{Hxc}$ in \eqref{dyson-approx} can be neglected. This leads to
\begin{equation}
\label{g0w0-equation}
   \left[\Sigma_{xc}\right]_{pp}(\epsilon_p) - \left[v_{xc}\right]_{pp} 
   = \epsilon_p - \epsilon^{KS}_p \;,
\end{equation}
Solving this equation as a perturbative correction is referred to as $G_0W_0$, while in ev$GW$, \cref{gw-self-energy-full,screened-coulomb-2,chi0,g0w0-equation} are solved self-consistently instead. Splitting the operator $\Sigma_{Hxc} - v_{Hxc}$ in \eqref{dyson-approx} into Hermitian and anti-Hermitian part and discarding the latter one, the solution of \eqref{dyson-approx} can be restricted to its QP part only.\cite{Layzer1963, Sham1966, Huser2013, Nakashima2021} Restricting the self-energy further to its static limit, a single-particle problem similar to the KS equations is obtained, 
\begin{equation}
\label{DysonqsGW}
    \sum_q\left\{\left[\Sigma^H_{Hxc}\right]_{pq} - \left[v_{Hxc} \right]_{pq}
    \right\}\phi_q(\br) = \left(\epsilon_p - \epsilon^{KS}_p\right) \phi_p(\br) \;,
\end{equation}
where $\Sigma^H = \frac{1}{2}\left(\Sigma + \Sigma^{\dagger}\right)$ denotes the Hermitian part of the self-energy. Solving \cref{gw-self-energy-full,screened-coulomb-2,chi0,DysonqsGW} self-consistently is referred to as the qs$GW$\cite{Faleev2004, VanSchilfgaarde2006, Kotani2007} approximation.\bibnote{It should be understood that in practice one solves 
\begin{equation}
\label{DysonqsGW2}
    \sum_q\left\{
    \left[\Sigma^H_{Hxc}\right]_{pq} - 
    \left[\Sigma^{H^{(n-1)}}_{Hxc}\right]_{pq}
    \right\}\phi_q(\br) = \left(\epsilon_p - \epsilon^{(n-1)}_p\right) \phi_p(\br) 
\end{equation}
in the $n$th iteration, which reduces to \eqref{DysonqsGW} for $n=1$.} There are many possible ways to construct the qs$GW$ Hamiltonian.\cite{Kotani2007, Shishkin2007, Kutepov2012, Kutepov2017b, Friedrich2022} In our implementation, we use the expression 
    \begin{equation}
    \label{ksf2}
        \left[\Sigma^{(GW)}\left(\left\{\epsilon_n\right\}\right)\right]_{pq}= 
        \begin{cases}
        \left[\Sigma^{(GW)}(\epsilon_p)\right]_{pq} & p = q \\ 
        \left[\Sigma^{(GW)}(\tilde{\epsilon})\right]_{pq} & \text{else} \;. 
        \end{cases}
    \end{equation}
with $\tilde{\epsilon} = 0$. If, as in our implementation\cite{Forster2021a}, the self-energy on the real frequency axis is calculated via analytical continuation (AC), eq.~\eqref{ksf2} is numerically more stable\cite{Forster2021a,Lei2022} than constructions of the qs$GW$ Hamiltonian in which also the off-diagonal elements are evaluated at the QP energies.\cite{Kotani2007,Kaplan2016}

\subsection{Kramers-restricted two-component formalism}
Recently, an 2C implementation of the GWA for Kramers-unrestricted systems has been implemented by Holzer with $\mathcal{O}\left( N^4\right)$ scaling with system size.\cite{Holzer2023} In this work we will focus on application to closed-shell molecules with no internal or external magnetic fields. This allows us to simplify the treatment considerably as it possible to define a Kramers-restricted set of spinors in which pairs of spinors are related by time-reversal symmetry.

We expand each molecular spinor in a primary basis of atomic orbitals (AO), $\left\{\chi_{\mu}\right\}_{\mu = 1, \dots, N_{\text{bas}}}$, as
\begin{equation}
\label{spinors}
 \phi_k(\br)  =   \left( \begin{array}{c} \phi^\uparrow_k(\br) \\ \phi^\downarrow_k(\br) \end{array} \right)
  = \sum_{\mu} \left( \begin{array}{c} b_{k \uparrow \mu} \chi_\mu(\br) \\ b_{k \downarrow \mu} \chi_\mu(\br) \end{array} \right) 
  = \sum_{\mu} \left( \begin{array}{c} (b_{k\uparrow\mu}^{R} + i b_{k\uparrow\mu }^{I}) \chi_\mu(\br) \\ (b_{k\downarrow\mu}^{ R} + i b_{k\downarrow \mu}^{I}) \chi_\mu(\br) \end{array} \right) \;,
\end{equation}
where $\uparrow$ ($\sigma=\frac{1}{2}$) and $\downarrow$ ($\sigma=-\frac{1}{2}$) denote the different projections of spin on the $z$-axis. Each spinor $\phi_{k}$ can be related by the time-reversal symmetry or Kramers' operator $\hat{K}$ to a Kramers' partner $\phi_{\bar{k}}$ with the same energy, $\epsilon_{k}=\epsilon_{\bar{k}}$, 
\begin{equation}
\label{kramers-symmetry}
    \hat{K} \phi_{k} =
    \begin{pmatrix}
    \phi_{k}^{\uparrow} (\br) \\ 
    \phi_{k}^{\downarrow} (\br) \\
    \end{pmatrix} = 
     \begin{pmatrix}
    -\phi_{k}^{\downarrow^*} (\br) \\ 
    \phi_{k}^{\uparrow*} (\br) \\
    \end{pmatrix}   = 
      \begin{pmatrix}
    -\phi_{k}^{\downarrow,R} (\br) 
    + i \phi_{k}^{\downarrow,I} (\br)  \\ 
    \phi_{k}^{\uparrow,R} (\br) 
    - i \phi_{k}^{\uparrow,I} (\br)  \\ 
    \end{pmatrix} 
    = \phi_{\bar{k}}\;.
\end{equation}
Using quaternion algebra it is possible to reduce the dimension of matrices that need to be considered to half the original size\cite{Saue1999}. Alternatively, one may keep the full dimension, but use the spinor pairing to define matrices as either real or imaginary. We will take the latter approach in this work. Denoting pairs of spinors with $\left(p, \bar{p} \right)$, noting that $\hat{K} \phi_{\bar{p}}=-\phi_p$ and transforming a purely imaginary diagonal operator $A$ that obeys $A_{pp}=A_{\bar{p}\bar{p}}$ and $A_{pp}=-A^*_{pp}$ we can deduce
\begin{equation}
\label{Kramersrelation}
\begin{aligned}
    A_{\mu\nu, \uparrow \uparrow} = & \sum_{p}  b_{p \uparrow \mu } A_{pp} b^*_{p \uparrow \nu} + 
    \sum_{\bar{p}} 
    b_{\bar{p} \uparrow \mu } A_{\bar{p}\bar{p}} b^*_{\bar{p} \uparrow \nu} = 
    \sum_{\bar{p}} 
    b^*_{\bar{p} \downarrow \mu } A_{\bar{p}\bar{p}}  b_{\bar{p} \downarrow \nu} + 
    \sum_{p} 
    b^*_{p \downarrow \mu } A_{pp} b_{p \downarrow \nu} = - A_{\mu\nu, \downarrow \downarrow}^* \\
    A_{\mu\nu, \downarrow \uparrow} = & \sum_{p}  b_{p \downarrow \mu } A_{pp} b^*_{p \uparrow \nu} + \sum_{\bar{p}} b_{\bar{p} \downarrow \mu } A_{\bar{p}\bar{p}} b^*_{\bar{p} \uparrow \nu} = 
    - \sum_{p}  
    b^*_{\bar{p} \uparrow \mu } 
    A_{\bar{p}\bar{p}} b_{\bar{p} \downarrow \nu} 
    - \sum_{\bar{p}} 
    b^*_{p \uparrow \mu } A_{pp} b_{p \downarrow \nu} = A_{\mu\nu, \uparrow \downarrow}^* \;.
\end{aligned}
\end{equation}
Is is convenient to split this operator into real and imaginary components, and we use the character of the MO coefficient products to label real (superscript R) and imaginary (superscript I) parts of the operator,  
\begin{equation}
   A^{R}_{\mu \nu, \sigma \sigma'} =  
   \sum_{p}  b^R_{p \sigma \mu } A_{pp} b^R_{p \sigma' \nu} + 
   \sum_{\bar{p} } b^R_{\bar{p} \sigma \mu } A_{\bar{p}\bar{p}} b^R_{\bar{p} \sigma' \nu} + 
   \sum_{p} b^I_{p \sigma \mu } A_{pp} b^I_{p \sigma' \nu} + 
   \sum_{\bar{p} } b^I_{\bar{p} \sigma \mu } A_{\bar{p}\bar{p}} b^I_{\bar{p} \sigma' \nu}
\end{equation}
and 
\begin{equation}
   A^{I}_{\mu \nu, \sigma \sigma'} =  
   \sum_{p}  b^R_{p \sigma \mu } A_{pp} b^I_{p \sigma' \nu} + 
   \sum_{\bar{p} } b^R_{\bar{p} \sigma \mu } A_{\bar{p}\bar{p}} b^I_{\bar{p} \sigma' \nu} - 
   \sum_{p} b^I_{p \sigma \mu } A_{pp} b^R_{p \sigma' \nu} - 
   \sum_{\bar{p} } b^I_{\bar{p} \sigma \mu } A_{\bar{p}\bar{p}} b^R_{\bar{p} \sigma' \nu} \;.
\end{equation}
The  time-ordered single-particle Green's function fulfills \cref{Kramersrelation} and therefore in AO basis obeys the relations
\begin{equation}
\label{greensKramers}
\begin{aligned}
    G^{\lessgtr}_{\mu \nu, \upuparrows}(i\tau) = & - G^{\lessgtr*}_{\mu \nu, \downdownarrows}(i\tau) \\
    G^{\lessgtr}_{\mu \nu, \updownarrows}(i\tau) = & G^{\lessgtr*}_{\mu \nu, \downuparrows}(i\tau) \;.
\end{aligned}
\end{equation}
Convenient is sometimes also to re-express these quantities in a spin matrix basis. We then get (denoting the
unit matrix as 0, and the Pauli spin matrices as x, y and z)
\begin{equation}
\label{greensGWbasisU}
\begin{aligned}
    G^{\lessgtr^0}_{\mu\nu}(i\tau) 
    = & G^{\lessgtr}_{\mu \nu, \upuparrows}(i\tau)  + G^{\lessgtr}_{\mu \nu, \downdownarrows} (i\tau)
    = & 2 G^{\lessgtr^R}_{\mu \nu, \upuparrows}(i\tau)  , \\
   G^{\lessgtr^x}_{\mu\nu}(i\tau) 
    = & G^{\lessgtr}_{\mu \nu, \updownarrows}(i\tau)  + G^{\lessgtr}_{\mu \nu, \downuparrows} (i\tau)
    = & 2 G^{\lessgtr^I}_{\mu \nu, \updownarrows}(i\tau)   ,\\
   G^{\lessgtr^y}_{\mu\nu}(i\tau) 
    = & i G^{\lessgtr}_{\mu \nu, \updownarrows}(i\tau)  - i G^{\lessgtr}_{\mu \nu, \downuparrows} (i\tau)
    = & 2i G^{\lessgtr^R}_{\mu \nu, \updownarrows}(i\tau)  ,\\
   G^{\lessgtr^z}_{\mu\nu}(i\tau) 
    = & G^{\lessgtr}_{\mu \nu, \upuparrows}(i\tau)  - G^{\lessgtr}_{\mu \nu, \downdownarrows} (i\tau) 
    = & 2 G^{\lessgtr^I}_{\mu \nu, \upuparrows}(i\tau)  \;,
\end{aligned}
\end{equation}
which more clearly shows the relation to 1-component theories in which only the first Green's function has a non-zero value.

\subsubsection{Polarizability in imaginary time} 
We next consider the polarizability\cite{Aryasetiawan2008,Aryasetiawan2009,Sakuma2011}. Whereas in the complete formalism of Aryasetiawan and Biermann\cite{Aryasetiawan2008} 
the polarizability includes the response of the charge density to magnetic fields as well as the induction of current densities, both of these are considered strictly zero in a Kramers-restricted formalism.  
We can then define the relevant part of the polarizability in AO basis as
\begin{equation}
\label{4terms}
P^{(0)}_{\mu\nu\sigma,\kappa\lambda\sigma'}(i\tau) = 
i\Theta(\tau) 
G^{>}_{\mu\kappa,\sigma\sigma'}(i\tau)
G^{<}_{\nu\lambda,\sigma'\sigma}(-i\tau)
+ i\Theta(-\tau) 
G^{<}_{\mu\kappa,\sigma\sigma'}(i\tau)
G^{>}_{\nu\lambda,\sigma'\sigma}(-i\tau) \;.
\end{equation} 
Due to the symmetry $P^{(0)}(i\tau) = P^{(0)}(-i\tau)$, we can focus on the first term which we split in terms of real (R) and imaginary (I) components
\begin{equation}
\begin{aligned}
G^>_{\mu \kappa, \sigma \sigma'}(i\tau)
G^<_{\nu \lambda, \sigma' \sigma}(-i\tau)
= & 
G^{>^R}_{\mu \kappa, \sigma \sigma'}(i\tau)
G^{<^R}_{\nu \lambda, \sigma' \sigma}(-i\tau) - 
G^{>^I}_{\mu \kappa, \sigma \sigma'}(i\tau)
G^{<^I}_{\nu \lambda, \sigma' \sigma}(-i\tau) \\ 
+ &
i G^{>^I}_{\mu \kappa, \sigma \sigma'}(i\tau)
G^{<^R}_{\nu \lambda, \sigma' \sigma}(-i\tau) + 
i G^{>^R}_{\mu \kappa, \sigma \sigma'}(i\tau)
G^{<^I}_{\nu \lambda, \sigma' \sigma}(-i\tau)  \;.
\end{aligned}
\end{equation}
Kramers symmetry implies
\begin{equation}
\label{kramers1}
\sum_{\sigma, \sigma' = \uparrow,\downarrow}
i G^{>^I}_{\mu \sigma, \kappa  \sigma'}(i\tau)
G^{<^R}_{\nu \sigma', \lambda \sigma}(-i\tau) + 
i G^{>^R}_{\mu \sigma,\kappa  \sigma'}(i\tau)
G^{<^I}_{\nu \sigma',  \lambda \sigma}(-i\tau) = 0 \;,
\end{equation} 
as well as 
\begin{equation}
\label{kramers2}
\begin{aligned}
P^{(0)}_{\mu\nu\uparrow,\kappa\lambda\uparrow}(i\tau) = & 
P^{(0)}_{\mu\nu\downarrow,\kappa\lambda\downarrow}(i\tau) \\
P^{(0)}_{\mu\nu\uparrow,\kappa\lambda\downarrow}(i\tau) = & 
P^{(0)}_{\mu\nu\downarrow,\kappa\lambda\uparrow}(i\tau) \;.
\end{aligned}
\end{equation}
We proof these relations in appendix~\ref{app::B}.
Already in the primary AO basis this would reduce the number of matrix elements that are to be calculated considerably. 
Further efficiency can be gained by expanding the polarizability and the Coulomb potential in a basis of auxiliary functions $\left\{f_{\alpha}\right\}_{\alpha = 1, \dots, N_{\text{aux}}}$
with products of primary basis functions being expressed as
\begin{equation}
\label{fitting}
    \chi_\mu(\br) \chi_\nu(\br) = \sum_{\alpha} c_{\mu \nu \alpha}  f_{\alpha}(\br) \;.
\end{equation}

To calculate the fitting coefficients, we use the pair-atomic density fitting (PADF) method\cite{Watson2003, Krykunov2009, Merlot2013, Merlot2014, Wirz2017, Ihrig2015} in the implementation of ref.~\citen{Spadetto2023}. The following working equations are however completely general and can be implemented using any type of density fitting (DF). For instance, global density fitting using the overlap kernel\cite{Dunlap1979} (also known as RI-SVS) or the attenuated Coulomb kernel\cite{Feyereisen1993, Jung2005} which have already been used to achieve low-scaling $GW$ implementations\cite{Wilhelm2018,Wilhelm2021} would be suitable choice as well.

For the polarizability we can eliminate the explicit dependence on spin in the transformation to the auxiliary basis and work with the spin-summed form

\begin{equation}
P^{(0)}_{\alpha\beta}(i\tau)  = 
\sum_{\sigma \sigma' = \uparrow, \downarrow}
c_{\mu \nu \alpha}  
P^{(0)}_{\mu\kappa\sigma,\nu\lambda\sigma'}(i\tau)
c_{\kappa \lambda \beta} \;.
\end{equation}
Likewise we define spin-independent representations of the Coulomb potential and screened interaction in the auxiliary basis as
\begin{align}
\label{pol_aux}
v_{\alpha \beta} = & \int d \br d \br' f_{\alpha}(\br)v_c(\br,\br')f_{\beta}(\br') \\
W_{\alpha \beta} (i\tau) = & \int d \br d \br' f_{\alpha}(\br)W(\br,\br',i\tau)f_{\beta}(\br') \;,
\end{align}
Our final expression for the polarizability is 
\begin{equation}
\label{final}
\begin{aligned}
P^{(0)}_{\alpha\beta}(i\tau) = -2i 
c_{\mu \nu \alpha}  & 
\left\{
G^{>^R}_{\mu\kappa,\uparrow\uparrow}(i\tau)
G^{<^R}_{\nu\lambda,\uparrow\uparrow}(i\tau) - 
G^{>^I}_{\mu\kappa,\uparrow\uparrow}(i\tau)
G^{<^I}_{\nu\lambda,\uparrow\uparrow}(i\tau)  \right. \\
& \left. \quad + 
G^{>^R}_{\mu\kappa , \uparrow\downarrow}(i\tau)
G^{<^R}_{\nu\lambda , \uparrow\downarrow}(i\tau) - 
G^{>^I}_{\mu\kappa,\uparrow\downarrow}(i\tau)
G^{<^I}_{\nu\lambda, \uparrow\downarrow}(i\tau)
\right\}
c_{\kappa \lambda \beta} \;,
\end{aligned}
\end{equation}
or equivalently 
\begin{equation}
\label{final_alternate}
\begin{aligned}
P^{(0)}_{\alpha\beta}(i\tau) = -\frac{1}{2}i 
c_{\mu \nu \alpha}  & 
\left\{
G^{>^0}_{\mu \kappa }(i\tau)
G^{<^0}_{\nu \lambda }(i\tau) - 
G^{>^x}_{\mu \kappa }(i\tau)
G^{<^x}_{\nu \lambda }(i\tau)  \right. \\
& \left. \quad - 
G^{>^y}_{\mu \kappa}(i\tau)
G^{<^y}_{\nu  \lambda}(i\tau) - 
G^{>^z}_{\mu \kappa}(i\tau)
G^{<^z}_{\nu \lambda}(i\tau) 
\right\}
c_{\kappa \lambda \beta} \;.
\end{aligned}
\end{equation}
The first term in this expression is equivalent in the spin-restricted 1C formalism.\cite{Forster2020b} Evaluation of \eqref{final} or \eqref{final_alternate} is therefore exactly four times more expensive than in a scalar relativistic calculation. \Cref{final} can be implemented with quadratic scaling with system size using PADF.\cite{Forster2020b}

\subsubsection{Polarizability in imaginary frequency and MO basis}
The AO based implementation of the polarizability is advantageous for rather large molecules only and it is computationally not efficient for the molecules in the SOC81 database typically containing just a few often heavy atoms. The AO based algorithms become advantageous when the local nature of the atomic orbitals can be exploited\cite{Forster2020b}. This is only possible when the system is spatially extended and many functions in the basis set decay fast with the distance from the nucleus on which they are centered. 

Especially for small systems with many heavy atoms, implementations in the canonical basis are much faster since in those systems the locality of the AO basis cannot be exploited. For this reason we also implemented the polarizability in the MO representation. In the following, we will use $i,j \dots$ to label occupied, and $a,b \dots$ to label virtual orbitals. Using \cref{g_time-ordered_mo} and these indices, \cref{4terms} becomes 
\begin{equation}
    P^{(0)}_{aiai}(i\tau) = -i\Theta(\tau) e^{-(\epsilon_a - \epsilon_i)\tau} - i\Theta(-\tau) e^{-(\epsilon_i - \epsilon_a)\tau} 
\end{equation}
in the MO basis. Using \eqref{TtoW}, the corresponding expression on the imaginary frequency axis is 
\begin{equation}
\label{P_mo}
    P^{(0)}_{aiai}(i\omega) =  - \frac{1}{\epsilon_a - \epsilon_i - i\omega} - \frac{1}{\epsilon_a - \epsilon_i + i\omega} \;.\\
\end{equation}
Using the last equation on the \emph{r.h.s.} of \eqref{spinors} and \eqref{fitting}, we can write down a transformation from the auxiliary basis to the MO basis as 
\begin{equation}
\label{transform1}
    \phi_i^\dagger(\br) \phi_a(\br) = \sum_{\alpha} c_{i a \alpha}  f_{\alpha}(\br) 
\end{equation}
with
\begin{equation}
\label{transform2}
    \begin{aligned}
         c_{i a \alpha}  = &  \sum_{\mu \kappa} (b_{i \uparrow \mu}^* b_{a \uparrow \kappa} + b_{i \downarrow \mu}^* b_{a  \downarrow \kappa} )c_{\mu \kappa \alpha}    = c^R_{i a \alpha} + i c^I_{i a \alpha} \\
    = & \sum_{\mu \kappa} (b_{i \uparrow \mu}^R b_{a \uparrow \kappa}^R + b_{i \uparrow \mu}^I b_{a \uparrow \kappa}^I + b_{i \downarrow \mu}^R b_{a \downarrow \kappa}^R + b_{i \downarrow \mu}^I b_{a \downarrow \kappa}^I ) c_{\mu \kappa \alpha}   \\
  & + i \sum_{\mu \kappa} (b_{i \uparrow \mu}^R b_{a \uparrow \kappa}^I - b_{i \uparrow \mu}^I b_{a \uparrow \kappa}^R + b_{i \downarrow \mu}^R b_{a \downarrow \kappa}^I - b_{i \downarrow \mu}^I b_{a \downarrow \kappa}^R ) c_{\mu \kappa \alpha} \;.
    \end{aligned}
\end{equation}
Using this expression, \cref{P_mo} becomes
\begin{equation}
\begin{aligned}
    P^{(0)}_{\alpha\beta}(i\omega) = & c_{ai\alpha}P^{(0)}_{aiai}(i\omega)c_{ai\beta} \\ 
    = & 
    2 \left\{c^R_{i a \alpha} \text{Re} P^{(0)}_{aiai} -  
    c^I_{i a \alpha} \text{Im} P^{(0)}_{aiai}\right\} c^R_{i a \beta} +
    2 \left\{c^R_{i a \alpha} \text{Im} P^{(0)}_{aiai} +  
    c^I_{i a \alpha} \text{Re} P^{(0)}_{aiai}\right\} c^I_{i a \beta} \;.    
\end{aligned}
\end{equation}

\subsubsection{Screened interaction and self-energy}
If necessary, the polarizability is transformed to the imaginary frequency axis where the screened interaction is calculated in the basis of auxiliary functions using \cref{screened-coulomb-2}, 
\begin{equation}
\label{wFitting}
W_{\alpha\beta}(i\omega) = v_{\alpha\beta} + \sum_{\gamma \delta}v_{\alpha \gamma} P^{(0)}_{\gamma \delta}(i\omega) W_{ \delta \gamma}(i\omega) \;.
\end{equation}
For the evaluation of the self-energy, we partition the screened Coulomb interaction as
\begin{equation}
    \widetilde{W} = W - v \;.
\end{equation}
This allows us to use different approximations for the dynamical and static contributions to the self-energy. To evaluate the self-energy on the imaginary frequency axis, we first define the time-ordered self-energy\cite{VanLeeuwen2015} 
\begin{equation}
\Sigma_{xc}(i\tau) = \Sigma_x + \Theta(\tau)\Sigma_c^{>}(i\tau) - \Theta(-\tau)\Sigma_c^{<}(i\tau) \;.
\end{equation} 
Here, the greater and lesser components of the self-energy are given by 
\begin{equation}
\label{sigma_c}
    \left[\Sigma_c^{\lessgtr}\right]_{\mu\nu,\sigma\sigma'}(i\tau) = 
    iG^{\lessgtr}_{\kappa \lambda, \sigma \sigma'}(i\tau)
c_{\mu \kappa \alpha}
\widetilde{W}_{\alpha\beta} (i\tau) c_{\nu \lambda \beta}\;,
\end{equation}
and the singular contribution (Fock term) as 
 \begin{equation}
\label{sigma_x}
    \left[\Sigma_x\right]_{\mu\nu,\sigma\sigma'} = 
    iG^{<}_{\kappa \lambda, \sigma \sigma'}(i\tau \rightarrow 0^-)
c_{\mu \kappa \alpha}
v_{\alpha\beta} c_{\nu \lambda \beta}\;.
\end{equation}   

\paragraph{Dynamical contribution}

In the basis of Pauli matrices, \eqref{sigma_c} can be expanded as
\begin{equation}
\label{sigma_c_expanded}
\left[\Sigma_c^{\lessgtr}\right]_{\mu\nu}(i\tau)  = 
i \left( \begin{array}{cc} 
 G^{\lessgtr^0}_{\kappa \lambda}(i\tau) +   G^{\lessgtr^z}_{\kappa \lambda}(i\tau) & G^{\lessgtr^x}_{\kappa \lambda}(i\tau) - i G^{\lessgtr^y}_{\kappa \lambda}(i\tau) \\
 G^{\lessgtr^x}_{\kappa \lambda}(i\tau) + i G^{\lessgtr^y}_{\kappa \lambda}(i\tau) & G^{\lessgtr^0}_{\kappa \lambda}(i\tau) -   G^{\lessgtr^z}_{\kappa \lambda}(i\tau)
  \end{array} \right)
c_{\mu \kappa \alpha}
\widetilde{W}_{\alpha\beta} (i\tau) c_{\nu \lambda \beta}.
\end{equation}

In the correlation part of the self-energy we only calculate the contribution due to $G^{\lessgtr^0}$, i.e., $G^{\lessgtr^x}$,$G^{\lessgtr^y}$, $G^{\lessgtr^z}$ are set to zero. Therefore, using \eqref{greensGWbasisU}, eq.~\eqref{sigma_c_expanded} reduces to 
\begin{equation}
\label{sigma_c_expanded_d0}
\left[\Sigma_c^{\lessgtr}\right]_{\mu\nu}(i\tau)  = 
2i \left( \begin{array}{cc} 
G^{\lessgtr^{R}}_{\kappa \lambda,\upuparrows}(i\tau) & 0\\
 0 & G^{\lessgtr^{R}}_{\kappa \lambda,\upuparrows}(i\tau) 
  \end{array} \right)
c_{\mu \kappa \alpha}
\widetilde{W}_{\alpha\beta} (i\tau) c_{\nu \lambda \beta} \;.
\end{equation}
This quantity has the form as in the 1C formalism and in the same way as in our 1C implementation.\cite{Forster2020b} Notice also, that $G^{\lessgtr^{R}}$ has a prefactor of $-i$ due to the definitions \cref{g_g_mo,g_l_mo}. We Fourier transform \eqref{sigma_c} to the imaginary frequency axis using \cref{TtoW}, for which we follow the treatment of Liu et al.\cite{Liu2016} From there, the self-energy is transformed back to the MO basis and analytically continued to real frequencies using the algorithm by Vidberg and Serene.\cite{Vidberg1977} For details on the AC for $G_0W_0$ and qs$GW$ we refer to our previous work.\cite{Forster2020b, Forster2021a}

\paragraph{Hartree-exchange contribution}
\Cref{sigma_x} is recovered from \eqref{sigma_c_expanded} by replacing $\widetilde{W}(i\tau)$ with $v_c$ and using $D = G^<(i\tau \rightarrow 0^-)$ instead of $G^<(i\tau)$. The resulting expression is identical to the ones typically implemented in 2C-Hartree--Fock codes,\cite{Armbruster2008, Desmarais2019}
\begin{equation}
\label{sigma_x_expanded}
\left[\Sigma_x\right]_{\mu\nu} = 
\left( \begin{array}{cc} 
 D^{^0}_{\kappa \lambda} +   
 D^{^z}_{\kappa \lambda} & 
 D^{^x}_{\kappa \lambda} 
 - i D^{^y}_{\kappa \lambda} \\
 D^{^x}_{\kappa \lambda}
 + i D^{^y}_{\kappa \lambda} & 
 D^{^0}_{\kappa \lambda} -   
 D^{^z}_{\kappa \lambda}
  \end{array} \right)
c_{\mu \kappa \alpha}
v_{\alpha\beta} c_{\nu \lambda \beta}
\end{equation}
where the different components of $D$ are obtained as the $i\tau \rightarrow 0$ limit of \cref{greensGWbasisU}.
In qs$GW$, we also need to evaluate the block-diagonal Hartree-contribution to the self-energy,
\begin{equation}
\label{sigma_H_expanded}
\left[\Sigma_H\right]_{\mu \nu} =    
\left( \begin{array}{cc} 
 D^{^0}_{\kappa \lambda}  & \\  & 
 D^{^0}_{\kappa \lambda} 
  \end{array} \right)
c_{\mu \nu \alpha}
v_{\alpha\beta} 
c_{\kappa \lambda \beta}
\end{equation}
The full qs$GW$ Hamiltonian is then constructed according to \cref{ksf2} and \cref{DysonqsGW2} is solved in the MO basis from the previous iteration. The new set of MO expansion coefficients and QP energies is then used to evaluate \cref{greensGWbasisU} in the next iteration.

\subsubsection{The $G3W2$ Correction}
As explained in ref.~\citen{Forster2022}, we evaluate the contribution of the $G3W2$ term to the self-energy as a perturbative correction to the solution of the GWA. Relying on the assumption that $GW$ already gives rather accurate QP energies we expand $\Sigma^{G3W2}$ around the $GW$ QP energies and obtain
\begin{equation}
\label{g3w2_qs}
    \epsilon^{GW + G3W2}_p = 
    \epsilon^{GW}_p + \Sigma_{pp}^{G3W2}(\epsilon^{GW}_{p}) \;,
\end{equation}
at zeroth order where $\Sigma_{pp}^{G3W2}$ is evaluated using the $GW$ QP energies obtained from the solution of \eqref{g0w0-equation} or \eqref{DysonqsGW}. We restrict ourselves to the statically screened $G3W2$ self-energy which is obtained from \eqref{g3w2-self-energy-full} by replacing both $W(1,2)$ with $W(1,2)\delta(t_1 - t_2)$.\cite{Forster2022} In terms of $G^{(s)}$ and in a basis of single-particle states (In case of $G_0W_0$ or ev$GW$ this would be the basis of KS states, in case of qs$GW$ the basis of qs$GW$ eigenstates), 
this term becomes\cite{Forster2020}
\begin{equation}
\label{SigmaG3W2-static}
\Sigma_{pp}^{G3W2}(\epsilon_{p}) = 
\sum^{occ}_{i}\sum^{virt}_{ab}
\frac{W(i\omega=0)_{paib} W(i\omega=0)_{aibp}}
{\epsilon_a + \epsilon_b  - \epsilon_i - \epsilon_{p}}   
- \sum^{occ}_{ij}\sum^{virt}_{a}
\frac{W(i\omega=0)_{piaj} W(i\omega=0)_{iajp}}
{\epsilon_a - \epsilon_i - \epsilon_j + \epsilon_{p}} \;,
\end{equation}
with
\begin{equation}
\label{screened stuff}
    W(i\omega=0)_{pqrs} = \int d\br d\br' 
    \phi_p(\br)
    \phi^\dagger_q(\br)
    W(\br,\br',i\omega=0)
    \phi_r(\br')
    \phi^\dagger_s(\br') \;.
\end{equation}
Using the transformation \cref{transform1,transform2} we write \eqref{screened stuff} as 
\begin{equation}
W(i\omega = 0)_{pqrs} = \sum_{\alpha} d_{pq\alpha}
c_{rs \beta} \;,
\end{equation}
with
\begin{equation}
d_{pq\alpha} = 
\sum_{\beta}
c_{pq \beta} 
W(i\omega = 0)_{\alpha \beta} \;.
\end{equation}
When complex matrix algebra is used, inserting this transformation into \eqref{SigmaG3W2-static} increases the computational effort by a factor of 16 (notice that the denominator is always real) compared to the 1C case. To reduce the computational effort, we use real matrix algebra and define the intermediates
\begin{equation}
\label{intermediates}
\begin{aligned}
W^{R/I,R/I}_{pqrs} = &
    \sum_{\alpha} d^{R/I}_{pq\alpha}
c^{R/I}_{rs \beta} \\
e_{pqrs} = & W^{R,R}_{pqrs}
- W^{I,I}_{pqrs} \\
f_{pqrs} = & W^{R,I}_{pqrs}
+ W^{I,R}_{pqrs} \;.
\end{aligned}
\end{equation}
The final self-energy correction \eqref{SigmaG3W2-static} is then evaluated as 
\begin{equation}
\label{SigmaG3W2-static_imp}
\Sigma_{pp}^{G3W2}(\epsilon_{p}) = 
\sum^{occ}_{i}\sum^{virt}_{ab}
\frac{e_{paib} e_{aibp} - f_{paib} f_{aibp}}
{\epsilon_a + \epsilon_b  - \epsilon_i - \epsilon_{p}}   
- \sum^{occ}_{ij}\sum^{virt}_{a}
\frac{e_{piaj} e_{iajp} - f_{piaj} f_{iajp}}
{\epsilon_a - \epsilon_i - \epsilon_j + \epsilon_{p}} \;.
\end{equation}
Here, the by far most expensive step is the calculation of the first four intermediates defined in the first equation of \eqref{intermediates}. Therefore, evaluating \eqref{SigmaG3W2-static_imp} is four times more expensive than the corresponding 1C expression.

\section{\label{sec::computationalDetails}Computational Details}
\subsection{Choice of 2C-Hamiltonian}
The 2C $GW$ equations have been implemented in a locally modified development version of the Slater Type orbital (STO) based ADF engine\cite{adf2022} within the Amsterdam modeling suite (AMS2022).\cite{Ruger2022} In principle, the implementation is independent of the choice of the particular choice of the 2C Hamiltonian. In the work, we use the zeroth-order regular approximation (ZORA) Hamiltonian by van Lenthe et al,\cite{VanLenthe1993, VanLenthe1994, VanLenthe1996} which can be written as\cite{VanLenthe1996} 
\begin{equation}
\label{full2C}
\hat{h}_1^{ZORA}(\br) = \hat{h}_1^{ZORA,SR} (\br)+\hat{h}_{1}^{ZORA,SO}(\br) \;.
\end{equation}
The first term,
\begin{equation}
\label{zora-sr}
\hat{h}_1^{ZORA,SR}(\br) = v_{ext}(\br)+ \vec{p} \frac{c^2}{2c^2 - v_{ext}(\br)} \vec{p} 
\end{equation}
describes scalar relativistic effects and we use this Hamiltonian in all 1C calculations. The second term 
\begin{equation}
\label{zora-so}
\hat{h}_{1}^{ZORA,SO}(\br) = \frac{c^2}{\left(2c^2 - v_{ext}(\br)\right)^2}   
\vec{\sigma} 
\cdot 
\left(\nabla v_{ext}(\br) 
\times \vec{p} \right) 
\end{equation}
accounts for SOC. We employ the Hamiltonian \eqref{full2C} in all of the following 2C calculations. We also tested two Hamiltonians obtained from an exact transformation of the 4-component Dirac equation to 2-components (X2C and RA-X2C, respectively. In the latter variant, a regular approach to calculate the transformation matrix is used).\cite{Dyall1997, Kutzelnigg2005} In the X2C and RA-X2C method implemented in ADF, first the 4-component Dirac equation for a model potential (MAPA) of the molecule is calculated for the given basis set, using the modified Dirac equation (MDE) by Dyall\cite{Dyall1994} for X2C, or using the regular approach\cite{Sadlej1994} to the modified Dirac equation (RA-MDE) for RA-X2C. In the basis set limit the MDE and the RA-MDE should yield same results for the model potential (MAPA) but using a finite basis set, the results for MDE and RA-MDE will differ.\cite{Visscher1999} In a next step, these 4-component equations are transformed to a 2C form.\cite{VanLenthe1996b} We found, that the particular choice of 2C Hamiltonian (ZORA, X2C or RA-X2C) only affects the final ionization potentials (IP) by a few 10 meV.

\subsection{Basis Sets}
In all calculations, we expand the spinors in \eqref{spinors} in all-electron STO basis sets of triple- and quadruple-$\zeta$ quality (TZ3P and QZ6P, respectively).\cite{Forster2021} The STO type basis sets in ADF are restricted to a maximum angular momentum of $l=3$, which complicates reaching the basis set limit for individual QP energies.\cite{Bruneval2020, Stuke2020} This is especially true for heavier elements with occupied $d$- or $f$-shells where higher angular momenta functions are needed to polarize the basis.\cite{Jensen2013} 

The numerical atomic orbital (NAO) based BAND engine\cite{TeVelde1991, Philipsen2022} of AMS can be used with basis functions of arbitrary angular momenta. To obtain converged QP energies we therefore augment our TZ3P and QZ6P basis sets with higher angular momenta functions and calculate scalar relativistic QP energies. In the choice of the higher angular momenta functions we follow the construction of the Sapporo-DKH3-(T,Q)ZP-2012 basis sets\cite{Noro2012,Noro2013} for all elements in the fourth to the sixth row of the periodic table. In the following we denote these basis sets as TZ3P+ and QZ6P+. Except for the Lanthanides, where the highest angular momenta are $l=5$ and $l=6$, the augmented TZ (QZ) basis set typically contains basis functions with angular momentum up to $l=4$ ($l=5$) for elements beyond the third row. The basis set definitions are included in the supporting information.

To calculate our final QP energies we first calculate complete basis set (CBS) limit extrapolated scalar relativistic QP energies with the BAND code using the expression
\begin{equation}
    \label{cbsExtra}
    \epsilon_n^{GW, \text{scalar}}(CBS) = \epsilon_n^{GW, \text{scalar}}(QZ6P+) - 
    \frac{\epsilon_n^{GW, \text{scalar}}(QZ6P+) - \epsilon_n^{GW, \text{scalar}}(TZ3P+)}{1 - \frac{N^{QZ}_{bas}}{N^{TZ}_{bas}}} \;,
\end{equation}
where $\epsilon_n^{GW, \text{scalar}}(QZ6P+)$ ($\epsilon_n^{GW, \text{scalar}}(TZ3P+)$) denotes the value of the QP energy calculated with QZ6P+ (TZ3P+) and $N^{QZ}_{bas}$ and $N^{TZ}_{bas}$ denote the respective numbers of basis functions (in spherical harmonics so that there are e.g. 5 $d$ and 7 $f$ functions). This expression is commonly used for the extrapolation of $GW$ QP energies to the complete basis set limit for localized basis functions.\cite{VanSetten2015} Spin-orbit corrections $\Delta_n^{2C}$ are then calculated with ADF using the QZ6P basis set,
\begin{equation}
   \Delta_n^{2C}(QZ6P) = \epsilon_n^{GW, \text{scalar}}(QZ6P) -  \epsilon_n^{GW, \text{2C}}(QZ6P)
\end{equation}
The corresponding QP energies are then obtained as 
\begin{align}
\label{gw_2C}
   \epsilon_n^{GW, \text{2C}}(CBS) = & \epsilon_n^{GW, \text{scalar}}(CBS) + \Delta_n^{2C}(QZ6P) \\
\label{g3w2_2C}
   \epsilon_n^{GW + G3W2, \text{2C}}(CBS) = & \epsilon_n^{GW, \text{2C}}(CBS) + \Sigma^{G3W2}_{nn}(QZ6P) \;.
\end{align}
This choice is well justified since the major part correction to the KS QP energies comes from the scalar relativistic part of the $GW$ correction. The spin-orbit correction and the $G3W2$ corrections are typically of the order of only a few hundred meV in magnitude (also see explicit values in the supporting information). Therefore, even relatively large errors in these quantities while only have a minor effect on the final results.  

\subsection{Technical Details}
We perform $G_0W_0$ calculations using PBE, PBE0 and BHLYP\cite{Becke1993} orbitals and eigenvalues. The latter functional contains 50 \% of exact exchange which is typically the optimal fraction for $G_0W_0$ QP energies for organic molecules.\cite{Bruneval2013, Bruneval2015, Zhang2022} ev$GW$ and qs$GW$ calculations are performed starting from PBE0 orbitals and eigenvalues. In all calculations we set the numerical quality to \emph{VeryGood}.\cite{Forster2020} The auxiliary bases used to expand 4-point correlation functions are automatically generated from products of primary basis functions. For this, we use a variant of an algorithm introduced in ref.~\citen{Ihrig2015} which has recently been implemented in ADF and BAND.\cite{Spadetto2023} The size of the auxiliary basis in this approach can be tuned by a single threshold which we set to $\epsilon_{aux}=1 \times 10^{-10}$ in all partially self-consistent calculations and to $\epsilon_{aux}=1 \times 10^{-8}$ for $G_0W_0$. This corresponds to a very large auxiliary basis which is typically around 12 times larger than the primary basis and eliminates PADF errors for relative energies of medium molecules almost completely.\cite{Spadetto2023} 

Imaginary time and imaginary frequency variables are discretized using non-uniform bases $\mathcal{T} = \left\{\tau_{\alpha}\right\}_{\alpha = 1, \dots N_{\tau}}$ and $\mathcal{W} = \left\{\omega_{\alpha}\right\}_{\alpha = 1, \dots N_{\omega}}$ of sizes $N_{\tau}$ and $N_{\omega}$, respectively, tailored to each system. More precisely, \eqref{TtoW} is implemented as
\begin{eqnarray}
    \overline{F}(i\omega_{\alpha}) = &  \Omega^{(c)}_{\alpha\beta} \overline{F}(i\tau_\beta) \\
    \underline{F}(i\omega_{\alpha}) = &  \Omega^{(s)}_{\alpha\beta} \underline{F}(i\tau_\beta) \;, 
\end{eqnarray}
where $\overline{F}$ and $\underline{F}$ denote even and odd parts of $F$. The transformation from imaginary frequency to imaginary time only requires the (pseudo)inversion of $\Omega^{(c)}$ and $\Omega^{(s)}$, respectively. Our procedure to calculate $\Omega^{(c)}$ and $\Omega^{(s)}$  as well as $\mathcal{T}$ and $\mathcal{W}$ follows Kresse and coworkers.\cite{Kaltak2014,Kaltak2014a,Liu2016} The technical specifications of our implementation have been described in the appendix of ref.~\citen{Forster2021}.

\subsection{\label{sec::computationalDetails-convergence}Convergence acceleration}

For the molecules in the SOC81 set, we have found that the ev$GW$ and ev$GW_0$ calculations converge within 5-8 iterations within an accuracy of a few meV when the DIIS implementation of ref.~\citen{Veril2018} is used. All ev$GW$ results presented in this work have been obtained using this DIIS implementation with a convergence criterion of 3 meV.

On the other hand, using our own DIIS implementation of ref.~\citen{Forster2021a} the qs$GW$ equations often do not convergence for the systems in the SOC81 set. As discussed in the literature,\cite{Forster2021,Monino2022} this issue is related to multiple QP solutions which seem to occur frequently in systems containing heavy elements. More sophisticated DIIS algorithms might offer a solution to this problem.\cite{Pokhilko2022} In addition to the switching between the QP peaks there is additional numerical strain which most likely arises from precision issues from the AC of the self-energy. Especially problematic are the off-diagonal elements of the self-energy matrix which should be zero at convergence. For a more detailed discussion we refer to our previous work.\cite{Forster2021a}

In this work, we have found a linear mixing strategy with adaptive mixing parameter $\alpha_{mix}$ to lead to stable convergence of the qs$GW$ SCF procedure after typically around 15 iterations. Specifically, we start the self-consistency cycle with $\alpha^{(0)}_{mix} = 0.3$. In case the SCF error decreases, we use the mixing parameter $\alpha^{(n)}_{mix} = \max \left\{1.2 \times \alpha^{(n-1)}_{mix},0.5\right\}$ in the $n$th iteration. In case the SCF error increases, we reset the mixing parameter to $\alpha^{(0)}_{mix}$.

\section{\label{sec::results}Results}

\subsection{\label{sec::results_west}Comparison to WEST}

In this section~\ref{sec::results_west}, we compare our results for SOC81 to the ones calculated by Scherpelz et al.\cite{Scherpelz2016} with the WEST code.\cite{Govoni2015, Yu2022}

\subsubsection{Multi-solution Cases}

Before discussing the results in detail, we notice that Scherpelz and Govoni identified in total 14 systems\bibnote{In principle, there are 15 systems with multiple solutions. However, for \ce{CI4}, all three solutions are very close to each other. Therefore, we retain this system in our benchmark.} in the SOC81 set for which the non-linear QP equations \eqref{g0w0-equation} have multiple solutions for $G_0W_0$@PBE.\cite{Scherpelz2016} All of these solutions can be found graphically in the sum-over-states formalism (analytical integration of the self-energy)\cite{Bruneval2012, VanSetten2013, Bintrim2021} or contour deformation techniques,\cite{Lebegue2003, Govoni2015, Scherpelz2016, Golze2018} by plotting the self-energy matrix elements as a function of frequency. Also in cases where QP spectra are calculated with different basis sets it is possible to identify the matching peaks in individual spectra and perform a reliable extrapolation to the CBS limit.

AC, however, typically fails to detect all solutions in these cases. Furthermore, the resulting QP energies will be rather inaccurate since it is impossible to build a Padé model which reliably represents the energy dependence of self-energy matrix elements with strongly varying frequency dependence (see ref.~\citen{Scherpelz2016} for examples).\cite{Golze2018,Duchemin2020}

The occurrence of multiple solutions can be an artefact of the starting point used in a $G_0W_0$ calculation.\cite{Govoni2018} It can also caused by a break-down of the single QP picture caused by pronounced static correlation effects. The occurrence of multiple solutions complicates the comparison of our results to WEST, since it is not clear if the same solutions are compared. It also complicates the extrapolation of results to the CBS limit since it is unclear if the same QP solution is found for all basis sets. Also comparison to experimental data is difficult since it is unclear if the detected solutions correspond to QP or to satellite peaks in the experimental spectra. For all these reasons, we decided to exclude these systems from the following benchmark. This leaves us with 67 systems to which we refer to as SOC81*. 

\subsubsection{Scalar relativistic Ionization potentials}

\renewcommand*{\arraystretch}{0.4}
\sisetup{
  round-mode          = places, 
  round-precision     = 2, 
}

\noindent\begin{longtable}[c]{l
S[table-format=3.2]%
S[table-format=3.2]%
S[table-format=3.2]%
S[table-format=3.2]%
S[table-format=3.2]%
S[table-format=3.2]%
S[table-format=3.2]%
S[table-format=3.2]%
}
\caption{\label{tab::IPs_WEST}Scalar relativistic and 2C $G_0W_0$@PBE and $G_0W_0$@PBE0 ionization potentials (IP) for the SOC81* database calculated with ADF/BAND. The corresponding values from WEST are given for comparison. All values are in eV.} \\
\toprule
&  \multicolumn{4}{c}{ADF/BAND} & \multicolumn{4}{c}{WEST} \\
&  \multicolumn{2}{c}{scalar} & \multicolumn{2}{c}{2C} & 
\multicolumn{2}{c}{scalar} & \multicolumn{2}{c}{2C} \\ 
\cline{2-9}
 &  {$G_0W_0$@} & 
 {$G_0W_0$@} &
 {$G_0W_0$@} &
 {$G_0W_0$@} &
 {$G_0W_0$@} &
 {$G_0W_0$@} &
 {$G_0W_0$@} &
 {$G_0W_0$@} \\
Name & {PBE} & {PBE0} & {PBE} & {PBE0} & {PBE} & {PBE0} & {PBE} & {PBE0} \\
\midrule\endfirsthead\toprule 
&  \multicolumn{4}{c}{ADF/BAND} & \multicolumn{4}{c}{WEST} \\
&  \multicolumn{2}{c}{scalar} & \multicolumn{2}{c}{2C} & 
\multicolumn{2}{c}{scalar} & \multicolumn{2}{c}{2C} \\ 
\cline{2-9}
 &  {$G_0W_0$@} & 
 {$G_0W_0$@} &
 {$G_0W_0$@} &
 {$G_0W_0$@} &
 {$G_0W_0$@} &
 {$G_0W_0$@} &
 {$G_0W_0$@} &
 {$G_0W_0$@} \\
Name & {PBE} & {PBE0} & {PBE} & {PBE0} & {PBE} & {PBE0} & {PBE} & {PBE0} \\
\midrule\endhead\bottomrule\midrule%
\multicolumn{9}{r}{{Continued on next page}} \\ \bottomrule
\endfoot\bottomrule\endlastfoot
  {\ce{Al2Br6}} &  10.32 &  10.73 &  10.30 &  10.70 &  10.38 &  10.78 &  10.34 &  10.74 \\
   {\ce{AlBr3}} &  10.47 &  10.85 &  10.44 &  10.81 &  10.53 &  10.91 &  10.48 &  10.86 \\
    {\ce{AlI3}} &   9.32 &   9.67 &   9.19 &   9.53 &   9.44 &   9.78 &   9.23 &   9.57 \\
   {\ce{AsBr3}} &   9.79 &  10.14 &   9.76 &  10.09 &   9.83 &  10.17 &   9.77 &  10.10 \\
   {\ce{AsCl3}} &  10.53 &  10.89 &  10.53 &  10.88 &  10.67 &  11.00 &  10.66 &  10.99 \\
    {\ce{AsF3}} &  12.38 &  12.80 &  12.38 &  12.80 &  12.49 &  12.89 &  12.49 &  12.89 \\
    {\ce{AsF5}} &  14.48 &  15.31 &  14.47 &  15.30 &  14.51 &  15.28 &  14.49 &  15.26 \\
    {\ce{AsH3}} &  10.42 &  10.54 &  10.42 &  10.54 &  10.33 &  10.55 &  10.33 &  10.54 \\
    {\ce{AsI3}} &   8.86 &   9.33 &   8.70 &   9.09 &   8.99 &   9.39 &   8.72 &   9.11 \\
     {\ce{Br2}} &  10.29 &  10.55 &  10.16 &  10.40 &  10.33 &  10.59 &  10.16 &  10.42 \\
    {\ce{BrCl}} &  10.69 &  10.98 &  10.59 &  10.87 &  10.79 &  11.06 &  10.67 &  10.93 \\
{\ce{C10H10Ru}} &   6.89 &   8.72 &   6.89 &   8.72 &   7.00 &   {--} &   6.90 &   {--} \\
  {\ce{C2H2Se}} &   8.48 &   8.72 &   8.47 &   8.72 &   8.48 &   8.72 &   8.48 &   8.72 \\
  {\ce{C2H6Cd}} &   8.86 &   9.16 &   8.86 &   9.16 &   8.82 &   9.09 &   8.83 &   9.10 \\
  {\ce{C2H6Hg}} &   9.10 &   9.30 &   9.12 &   9.33 &   8.99 &   9.20 &   9.07 &   9.28 \\
  {\ce{C2H6Se}} &   8.14 &   8.38 &   8.14 &   8.38 &   8.17 &   8.41 &   8.12 &   8.41 \\
  {\ce{C2H6Zn}} &   9.42 &   9.70 &   9.42 &   9.70 &   9.38 &   9.65 &   9.38 &   9.65 \\
  {\ce{C2HBrO}} &   9.04 &   9.36 &   9.03 &   9.35 &   8.98 &   9.28 &   8.97 &   9.27 \\
  {\ce{C4H4Se}} &   8.72 &   8.98 &   8.72 &   8.98 &   8.60 &   8.86 &   8.60 &   8.86 \\
    {\ce{CF3I}} &  10.37 &  10.66 &  10.12 &  10.39 &  10.52 &  10.81 &  10.20 &  10.48 \\
 {\ce{CH3HgBr}} &   9.56 &   9.98 &   9.48 &   9.87 &   9.72 &  10.11 &   9.59 &   9.97 \\
 {\ce{CH3HgCl}} &  10.09 &  10.64 &  10.07 &  10.61 &  10.30 &  10.76 &  10.26 &  10.72 \\
  {\ce{CH3HgI}} &   8.85 &   9.23 &   8.66 &   9.00 &   9.08 &   9.38 &   8.79 &   9.09 \\
    {\ce{CH3I}} &   9.42 &   9.63 &   9.19 &   9.36 &   9.57 &   9.78 &   9.26 &   9.46 \\
     {\ce{CI4}} &   8.86 &   9.26 &   8.64 &   9.05 &   8.91 &   9.28 &   8.76 &   9.04 \\
   {\ce{CaBr2}} &   9.66 &  10.10 &   9.58 &   9.99 &   9.80 &  10.24 &   9.67 &  10.10 \\
    {\ce{CaI2}} &   8.99 &   9.29 &   8.79 &   9.06 &   9.09 &   9.48 &   8.79 &   9.18 \\
   {\ce{CdBr2}} &  10.05 &  10.49 &   9.95 &  10.36 &  10.21 &  10.62 &  10.06 &  10.46 \\
   {\ce{CdCl2}} &  10.72 &  11.23 &  10.70 &  11.19 &  10.91 &  11.39 &  10.87 &  11.34 \\
    {\ce{CdI2}} &   9.29 &   9.62 &   9.06 &   9.36 &   9.41 &   9.76 &   9.09 &   9.43 \\
     {\ce{CsF}} &   8.49 &   9.51 &   8.49 &   9.50 &   8.24 &   9.08 &   8.24 &   9.08 \\
   {\ce{HgCl2}} &  10.66 &  11.08 &  10.61 &  11.02 &  10.93 &  11.35 &  10.85 &  11.28 \\
      {\ce{I2}} &   9.34 &   9.64 &   9.05 &   9.34 &   9.41 &   9.64 &   9.01 &   9.26 \\
     {\ce{IBr}} &   9.70 &   9.95 &   9.45 &   9.69 &   9.81 &  10.04 &   9.51 &   9.73 \\
     {\ce{ICl}} &   9.99 &  10.24 &   9.74 &   9.97 &  10.14 &  10.37 &   9.85 &  10.07 \\
      {\ce{IF}} &  10.44 &  10.66 &  10.14 &  10.34 &  10.55 &  10.80 &  10.23 &  10.47 \\
     {\ce{Kr2}} &  13.28 &  13.57 &  13.19 &  13.45 &  13.42 &  13.68 &  13.27 &  13.53 \\
    {\ce{KrF2}} &  12.56 &  13.28 &  12.50 &  13.22 &  12.58 &  13.30 &  12.51 &  13.22 \\
   {\ce{LaBr3}} &   9.80 &  10.32 &   9.77 &  10.24 &   9.90 &  10.41 &   9.82 &  10.31 \\
   {\ce{LaCl3}} &  10.58 &  11.17 &  10.57 &  11.15 &  10.73 &  11.26 &  10.72 &  11.24 \\
    {\ce{LiBr}} &   8.79 &   9.16 &   8.70 &   9.05 &   8.95 &   9.35 &   8.81 &   9.21 \\
     {\ce{LiI}} &   8.10 &   8.48 &   7.90 &   8.25 &   8.35 &   8.65 &   8.04 &   8.36 \\
   {\ce{MgBr2}} &  10.37 &  10.79 &  10.27 &  10.67 &  10.49 &  10.91 &  10.35 &  10.76 \\
    {\ce{MgI2}} &   9.52 &   9.87 &   9.30 &   9.62 &   9.62 &   9.97 &   9.31 &   9.65 \\
  {\ce{MoC6O6}} &   8.55 &  12.44 &   8.52 &   9.86 &   8.55 &   {--} &   8.50 &   {--} \\
    {\ce{OsO4}} &  11.82 &  12.44 &  11.82 &  12.42 &  11.74 &  12.41 &  11.74 &  12.41 \\
    {\ce{PBr3}} &   9.56 &   9.89 &   9.54 &   9.86 &   9.60 &   9.92 &   9.57 &   9.88 \\
   {\ce{POBr3}} &  10.57 &  11.05 &  10.51 &  10.93 &  10.55 &  11.01 &  10.49 &  10.93 \\
    {\ce{RuO4}} &  11.48 &  12.24 &  11.48 &  12.24 &  11.45 &  12.19 &  11.44 &  12.18 \\
   {\ce{SOBr2}} &  10.12 &  10.58 &   9.97 &  10.52 &  10.17 &  10.57 &  10.13 &  10.52 \\
   {\ce{SPBr3}} &   9.47 &   9.77 &   9.45 &   9.75 &   9.45 &   9.82 &   9.43 &   9.79 \\
   {\ce{SeCl2}} &   9.13 &   9.45 &   9.10 &   9.43 &   9.24 &   9.53 &   9.24 &   9.53 \\
    {\ce{SeO2}} &  11.05 &  11.65 &  11.04 &  11.64 &  11.03 &  11.61 &  11.03 &  11.60 \\
  {\ce{SiBrF3}} &  11.68 &  12.00 &  11.57 &  11.87 &  11.78 &  12.10 &  11.64 &  11.95 \\
   {\ce{SiH3I}} &   9.80 &  10.06 &   9.59 &   9.81 &   9.93 &  10.17 &   9.64 &   9.86 \\
   {\ce{SrBr2}} &   9.39 &   9.79 &   9.30 &   9.67 &   9.49 &   9.92 &   9.35 &   9.77 \\
   {\ce{SrCl2}} &   9.90 &  10.39 &   9.89 &  10.38 &  10.00 &  10.49 &   9.97 &  10.46 \\
    {\ce{SrI2}} &   8.79 &   9.07 &   8.60 &   8.84 &   8.82 &   9.21 &   8.51 &   8.90 \\
   {\ce{TiBr4}} &   9.93 &  10.54 &   9.85 &  10.46 &   9.98 &  10.57 &   9.89 &  10.47 \\
    {\ce{TiI4}} &   8.77 &   9.35 &   8.61 &   9.17 &   8.90 &   9.42 &   8.65 &   9.17 \\
   {\ce{ZnBr2}} &  10.39 &  10.81 &  10.29 &  10.68 &  10.52 &  10.90 &  10.37 &  10.75 \\
   {\ce{ZnCl2}} &  11.19 &  11.66 &  11.16 &  11.62 &  11.36 &  11.79 &  11.32 &  11.75 \\
    {\ce{ZnF2}} &  12.57 &  13.30 &  12.56 &  13.28 &  12.69 &  13.42 &  12.66 &  13.39 \\
    {\ce{ZnI2}} &   9.55 &   9.89 &   9.32 &   9.62 &   9.63 &   9.96 &   9.30 &   9.63 \\
   {\ce{ZrBr4}} &  10.20 &  10.75 &  10.15 &  10.67 &  10.26 &  10.78 &  10.17 &  10.68 \\
   {\ce{ZrCl4}} &  11.26 &  11.82 &  11.25 &  11.80 &  11.35 &  11.93 &  11.32 &  11.91 \\
    {\ce{ZrI4}} &   9.20 &   9.57 &   9.04 &   9.38 &   9.17 &   9.62 &   8.93 &   9.36 \\
\midrule 
MSD & -0.07 & -0.06 & -0.04 & -0.4  &&&& \\
MAD & 0.10  &  0.09 & 0.07  & 0.07  &&&& \\
MAX & 0.27  &  0.43 & 0.25  & 0.42  &&&& \\
\bottomrule 
\end{longtable} 

\begin{table}[hbt!]
    \centering
    \begin{tabular}{lll}
    \toprule 
    & WEST & ADF/BAND \\
    \midrule 
     Single-particle basis     &  Plane-wave & Slater type orbital \\
     All-electron              &  No         & Yes \\ 
     Frequency treatment       &  Contour deformation & Analytical continuation \\
     QP equations              &  Secant method & Bisection \\
     Relativistic Hamiltonian  &  2C-pseudopotentials & ZORA \\
     2C self-energy            &  Static part only & Static and Dynamic part \\
     \bottomrule
    \end{tabular}
    \caption{Comparison of the implementations of 2C-$G_0W_0$ in WEST and ADF/BAND.}
    \label{tab::technical_details}
\end{table}

Table~\ref{tab::IPs_WEST} shows our scalar relativistic and 2C IPs using $G_0W_0$@PBE and $G_0W_0$@PBE0 and for comparison the corresponding values from ref.~\citen{Scherpelz2016} calculated with the WEST code. As indicated by the mean signed deviations (MSD) in table~\ref{tab::IPs_WEST}, ADF/BAND tends to predict lower IPs than WEST, independent of the starting point of the $G_0W_0$ calculation. With mean absolute deviations (MAD) of 100 meV for $G_0W_0$@PBE and 90 meV for $G_0W_0$@PBE0 in the scalar relativistic case and of 70 meV each in the 2C case, the deviations are of the same order of magnitude as the ones we obtained for the GW100 database.\cite{Govoni2018,Forster2021} 

Several technical aspects of the $GW$ implementations in ADF/BAND and WEST which are summarized in table~\ref{tab::technical_details} might contribute to the observed deviations. As discussed in the preceding section, these are mainly related to the different frequency treatments in both codes as well as differences in the single-particle basis. Importantly, WEST is based on PPs while we used all-electron basis sets in all ADF and BAND calculations. As already discussed extensively by Scherpelz and Govoni,\cite{Scherpelz2016} the choice of the PP and the partitioning of core, semi-core and valence electrons might heavily affect the values of the IPs. For instance, in ref.~\citen{Scherpelz2016}, it was shown that using different valence configurations for iodine might induce changes in IPs of the order of one eV. 

In all-electron calculations, this issue is completely avoided. However, possible issues might arise from inconsistencies in the augmentation of the TZ3P and QZ6P basis sets with additional high-$l$ functions. While it can be verified by comparison of TZ3P (QZ6P) results to their TZ3P+ ( QZ6P+) counterparts that adding any higher angular momenta functions will improve the quality of the AO basis, the effect is typically more pronounced on the TZ than on the QZ level. This might then lead to larger inaccuracies in the CBS limit extrapolation than in plane-wave based implementations.

\subsubsection{Changes in Ionization Potentials due to Spin-Orbit Coupling} 

Finally, the agreement between ADF/BAND and WEST is slightly better for the 2C than for the scalar relativistic calculations. This can be explained by the different division of scalar and spin–orbit relativistic effects in both codes (see table~\ref{tab::technical_details}). In particular, the division between scalar relativistic and SOC effects is not unique and depends on the method of separation.\cite{Visscher1999} 

\begin{figure}[hbt!]
    \centering
    \includegraphics[width=0.7\textwidth]{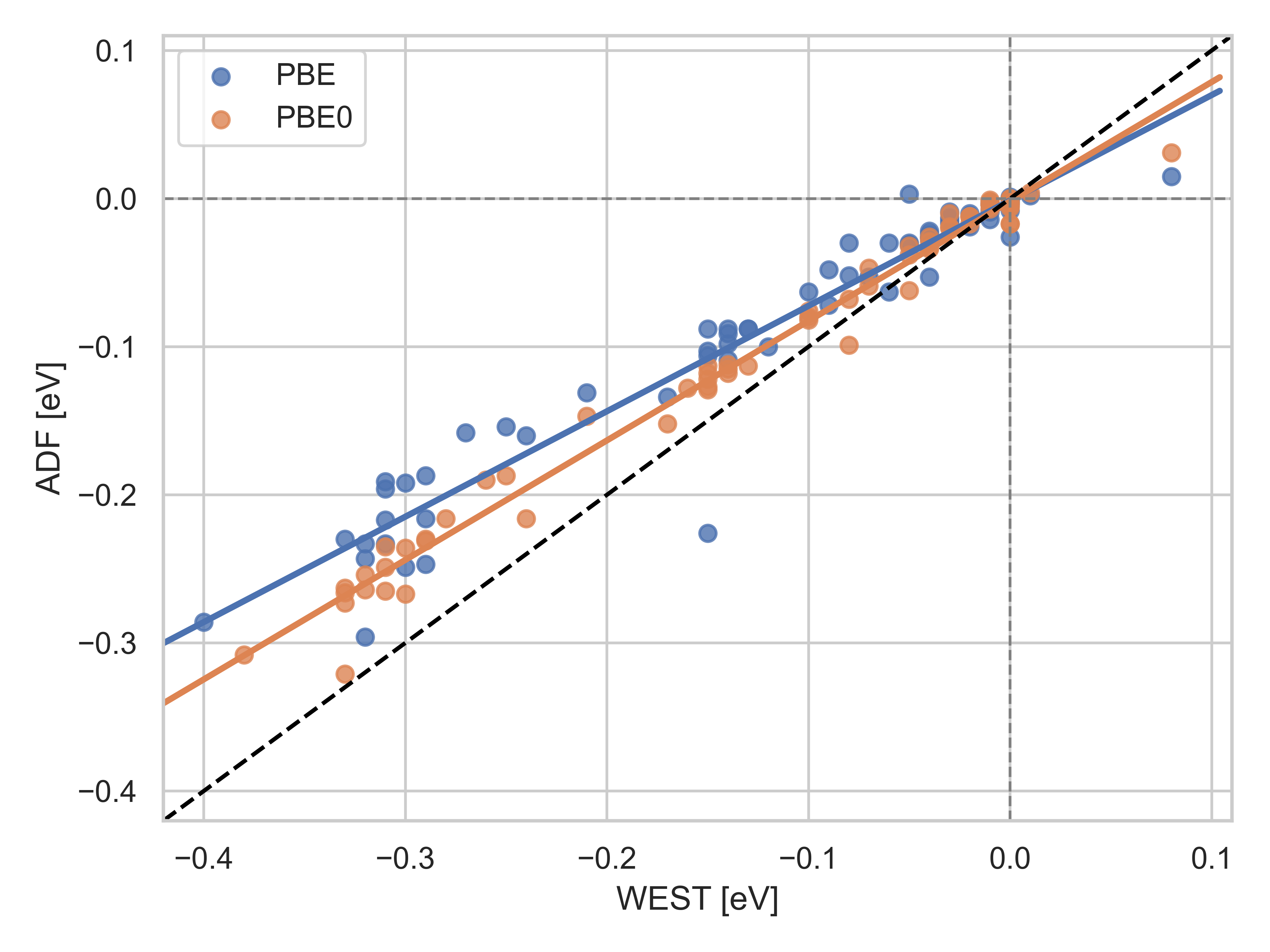}
    \caption{Comparison of the IP shift due to spin-orbit coupling as calculated with ADF compared to WEST for $G_0W_0$@PBE and $G_0W_0$@PBE0. All values are in eV.}
    \label{fig::so_splitting_west}
\end{figure}

This is also illustrated by the data shown in fig.~\ref{fig::so_splitting_west} where we plot the difference between the first IP in the scalar and the 2C relativistic case calculated with WEST (x-axis) against the one calculated with ADF. Overall, we find good agreement between both implementations. WEST tends to predict slightly larger shifts due to SO coupling than ADF, especially for $G_0W_0$@PBE. This most likely indicates that ADF/BAND recovers more of the relativistic effects in the scalar relativistic treatment than WEST. At the $G_0W_0$@PBE level we also notice one significant outlier (\ce{CI4}) where ADF/BAND predicts significantly larger shifts due to SO coupling than WEST.

\subsection{\label{sec::results_exp}Comparison to experiment}

\renewcommand*{\arraystretch}{0.4}
\sisetup{
  round-mode          = places, 
  round-precision     = 2, 
}

\noindent\begin{longtable}[c]{ll
S[table-format=3.2]%
S[table-format=3.2]%
S[table-format=3.2]%
S[table-format=3.2]%
S[table-format=3.2]%
S[table-format=3.2]%
S[table-format=3.2]%
}
\caption{\label{tab::IPs}First ionization potentials (IP) for the SOC81* database calculated with different 2C $GW$ methods. All values are in eV.} \\
\toprule
& \multicolumn{3}{c}{$G_0W_0$} & & & & \\ \cline{2-4} Name & {PBE} & {PBE0} & {BHLYP}  & {ev$GW_0$@PBE0} & {ev$GW$@PBE0} & {qs$GW$} & {exp.} \\
\midrule\endfirsthead\toprule 
& \multicolumn{3}{c}{$G_0W_0$} & & & & \\ \cline{2-4} Name & {PBE} & {PBE0} & {BHLYP}  
& {ev$GW_0$@PBE0} & {ev$GW$@PBE0} & {qs$GW$} & {exp.} \\
\midrule\endhead\bottomrule\midrule%
\multicolumn{8}{r}{{Continued on next page}} \\ \bottomrule
\endfoot\bottomrule\endlastfoot
  {\ce{Al2Br6}} &  10.30 &  10.70 &  10.98 &  10.92 &  11.09 &  11.24 &  10.97 \\ 
   {\ce{AlBr3}} &  10.44 &  10.81 &  11.06 &  11.03 &  11.19 &  11.31 &  10.91 \\ 
    {\ce{AlI3}} &   9.19 &   9.53 &   9.76 &   9.69 &   9.83 &   9.72 &   9.66 \\ 
   {\ce{AsBr3}} &   9.76 &  10.09 &  10.33 &  10.26 &  10.38 &  10.50 &  10.21 \\ 
   {\ce{AsCl3}} &  10.53 &  10.88 &  11.15 &  11.05 &  11.17 &  11.40 &  10.90 \\ 
    {\ce{AsF3}} &  12.38 &  12.80 &  13.14 &  13.03 &  13.21 &  13.46 &  13.00 \\ 
    {\ce{AsF5}} &  14.47 &  15.30 &  15.81 &  15.74 &  16.13 &  16.62 &  15.53 \\ 
    {\ce{AsH3}} &  10.42 &  10.54 &  10.70 &  10.70 &  10.78 &  10.79 &  10.58 \\ 
    {\ce{AsI3}} &   8.70 &   9.11 &   9.34 &   9.19 &   9.28 &   9.41 &   9.00 \\ 
     {\ce{Br2}} &  10.16 &  10.40 &  10.58 &  10.57 &  10.70 &  10.82 &  10.51 \\ 
    {\ce{BrCl}} &  10.59 &  10.87 &  11.06 &  11.04 &  11.17 &  11.33 &  11.01 \\ 
{\ce{C10H10Ru}} &   6.83 &   7.12 &   7.44 &   7.24 &   7.43 &   7.87 &   7.45 \\ 
  {\ce{C2H2Se}} &   8.47 &   8.72 &   8.88 &   8.86 &   8.96 &   9.03 &   8.71 \\ 
  {\ce{C2H6Cd}} &   8.86 &   9.16 &   9.34 &   9.32 &   9.45 &   9.58 &   8.80 \\ 
  {\ce{C2H6Hg}} &   9.12 &   9.33 &   9.57 &   9.54 &   9.63 &   9.77 &   9.32 \\ 
  {\ce{C2H6Se}} &   8.14 &   8.38 &   8.57 &   8.55 &   8.66 &   8.72 &   8.40 \\ 
  {\ce{C2H6Zn}} &   9.42 &   9.70 &   9.89 &   9.89 &  10.04 &  10.10 &   9.40 \\ 
  {\ce{C2HBrO}} &   9.03 &   9.35 &   9.59 &   9.50 &   9.62 &   9.73 &   9.10 \\ 
  {\ce{C4H4Se}} &   8.72 &   8.98 &   9.16 &   9.13 &   9.24 &   9.24 &   8.86 \\ 
    {\ce{CF3I}} &  10.12 &  10.39 &  10.67 &  10.53 &  10.63 &  10.64 &  10.45 \\ 
 {\ce{CH3HgBr}} &   9.48 &   9.87 &  10.08 &  10.10 &  10.29 &  10.39 &  10.16 \\ 
 {\ce{CH3HgCl}} &  10.07 &  10.61 &  10.89 &  10.93 &  11.10 &  11.32 &  10.84 \\ 
  {\ce{CH3HgI}} &   8.66 &   9.00 &   9.20 &   9.20 &   9.33 &   9.28 &   9.25 \\ 
    {\ce{CH3I}} &   9.19 &   9.36 &   9.53 &   9.51 &   9.62 &   9.51 &   9.52 \\ 
     {\ce{CI4}} &   8.64 &   9.05 &   9.31 &   9.19 &   9.32 &   9.27 &   9.10 \\ 
   {\ce{CaBr2}} &   9.58 &   9.99 &  10.21 &  10.21 &  10.39 &  10.48 &  10.35 \\ 
    {\ce{CaI2}} &   8.79 &   9.06 &   9.27 &   9.24 &   9.38 &   9.19 &   9.39 \\ 
   {\ce{CdBr2}} &   9.95 &  10.36 &  10.59 &  10.61 &  10.79 &  10.92 &  10.58 \\ 
   {\ce{CdCl2}} &  10.70 &  11.19 &  11.51 &  11.50 &  11.71 &  11.97 &  11.44 \\ 
    {\ce{CdI2}} &   9.06 &   9.36 &   9.57 &   9.54 &   9.69 &   9.61 &   9.57 \\ 
     {\ce{CsF}} &   8.49 &   9.50 &   9.79 &   9.91 &  10.32 &  10.60 &   9.68 \\ 
   {\ce{HgCl2}} &  10.61 &  11.02 &  11.30 &  11.28 &  11.48 &  11.85 &  11.50 \\ 
      {\ce{I2}} &   9.05 &   9.34 &   9.45 &   9.45 &   9.55 &   9.40 &   9.35 \\ 
     {\ce{IBr}} &   9.45 &   9.69 &   9.85 &   9.83 &   9.93 &  10.05 &   9.85 \\ 
     {\ce{ICl}} &   9.74 &   9.97 &  10.19 &  10.12 &  10.22 &  10.23 &  10.10 \\ 
      {\ce{IF}} &  10.14 &  10.34 &  10.56 &  10.48 &  10.60 &  10.57 &  10.62 \\ 
     {\ce{Kr2}} &  13.19 &  13.45 &  13.69 &  13.65 &  13.78 &  13.90 &  13.77 \\ 
    {\ce{KrF2}} &  12.50 &  13.22 &  13.89 &  13.62 &  13.99 &  14.37 &  13.34 \\ 
   {\ce{LaBr3}} &   9.77 &  10.24 &  10.51 &  10.47 &  10.67 &  10.80 &  10.68 \\ 
   {\ce{LaCl3}} &  10.57 &  11.15 &  11.50 &  11.42 &  11.64 &  11.98 &  11.29 \\ 
    {\ce{LiBr}} &   8.70 &   9.05 &   9.23 &   9.28 &   9.44 &   9.48 &   9.44 \\ 
     {\ce{LiI}} &   7.90 &   8.25 &   8.40 &   8.43 &   8.56 &   8.42 &   8.44 \\ 
   {\ce{MgBr2}} &  10.27 &  10.67 &  10.88 &  10.90 &  11.07 &  11.14 &  10.85 \\ 
    {\ce{MgI2}} &   9.30 &   9.62 &   9.80 &   9.79 &   9.93 &   9.77 &  10.50 \\ 
  {\ce{MoC6O6}} &   8.52 &   8.74 &   9.01 &   8.83 &   8.91 &   9.07 &   8.50 \\ 
    {\ce{OsO4}} &  11.82 &  12.42 &  12.83 &  12.71 &  12.97 &  12.97 &  12.35 \\ 
    {\ce{PBr3}} &   9.54 &   9.86 &  10.11 &  10.01 &  10.13 &  10.27 &   9.99 \\ 
   {\ce{POBr3}} &  10.51 &  10.95 &  11.24 &  11.14 &  11.31 &  11.49 &  11.03 \\ 
    {\ce{RuO4}} &  11.48 &  12.24 &  12.72 &  12.52 &  12.82 &  13.25 &  12.15 \\ 
   {\ce{SOBr2}} &  10.07 &  10.52 &  10.80 &  10.70 &  10.85 &  11.02 &  10.54 \\ 
   {\ce{SPBr3}} &   9.45 &   9.75 &  10.00 &   9.94 &  10.09 &  10.28 &   9.89 \\ 
   {\ce{SeCl2}} &   9.10 &   9.43 &   9.69 &   9.61 &   9.71 &  10.00 &   9.52 \\ 
    {\ce{SeO2}} &  11.04 &  11.64 &  12.00 &  11.93 &  12.19 &  12.49 &  11.76 \\ 
  {\ce{SiBrF3}} &  11.57 &  11.87 &  12.09 &  12.04 &  12.18 &  12.27 &  12.46 \\ 
   {\ce{SiH3I}} &   9.59 &   9.81 &   9.99 &   9.98 &  10.09 &  10.00 &   9.78 \\ 
   {\ce{SrBr2}} &   9.30 &   9.67 &   9.88 &   9.89 &  10.08 &  10.17 &   9.82 \\ 
   {\ce{SrCl2}} &   9.89 &  10.38 &  10.65 &  10.64 &  10.86 &  11.10 &  10.20 \\ 
    {\ce{SrI2}} &   8.60 &   8.84 &   9.02 &   9.01 &   9.15 &   9.01 &   9.01 \\ 
   {\ce{TiBr4}} &   9.85 &  10.46 &  10.80 &  10.70 &  10.87 &  11.06 &  10.59 \\ 
    {\ce{TiI4}} &   8.61 &   9.17 &   9.47 &   9.35 &   9.51 &   9.42 &   9.27 \\ 
   {\ce{ZnBr2}} &  10.29 &  10.68 &  10.90 &  10.92 &  11.09 &  11.24 &  10.90 \\ 
   {\ce{ZnCl2}} &  11.16 &  11.62 &  11.91 &  11.91 &  12.11 &  12.34 &  11.80 \\ 
    {\ce{ZnF2}} &  12.56 &  13.28 &  13.72 &  13.85 &  14.30 &  14.73 &  13.91 \\ 
    {\ce{ZnI2}} &   9.32 &   9.62 &   9.83 &   9.81 &   9.94 &   9.87 &   9.76 \\ 
   {\ce{ZrBr4}} &  10.15 &  10.67 &  10.99 &  10.90 &  11.09 &  11.24 &  10.86 \\ 
   {\ce{ZrCl4}} &  11.25 &  11.80 &  12.20 &  12.08 &  12.32 &  12.62 &  11.94 \\ 
    {\ce{ZrI4}} &   9.04 &   9.38 &   9.68 &   9.57 &   9.71 &   9.65 &   9.55 \\ 
\bottomrule 
\end{longtable} 

\begin{figure}[hbt!]
    \centering
    \includegraphics[width=0.7\textwidth]{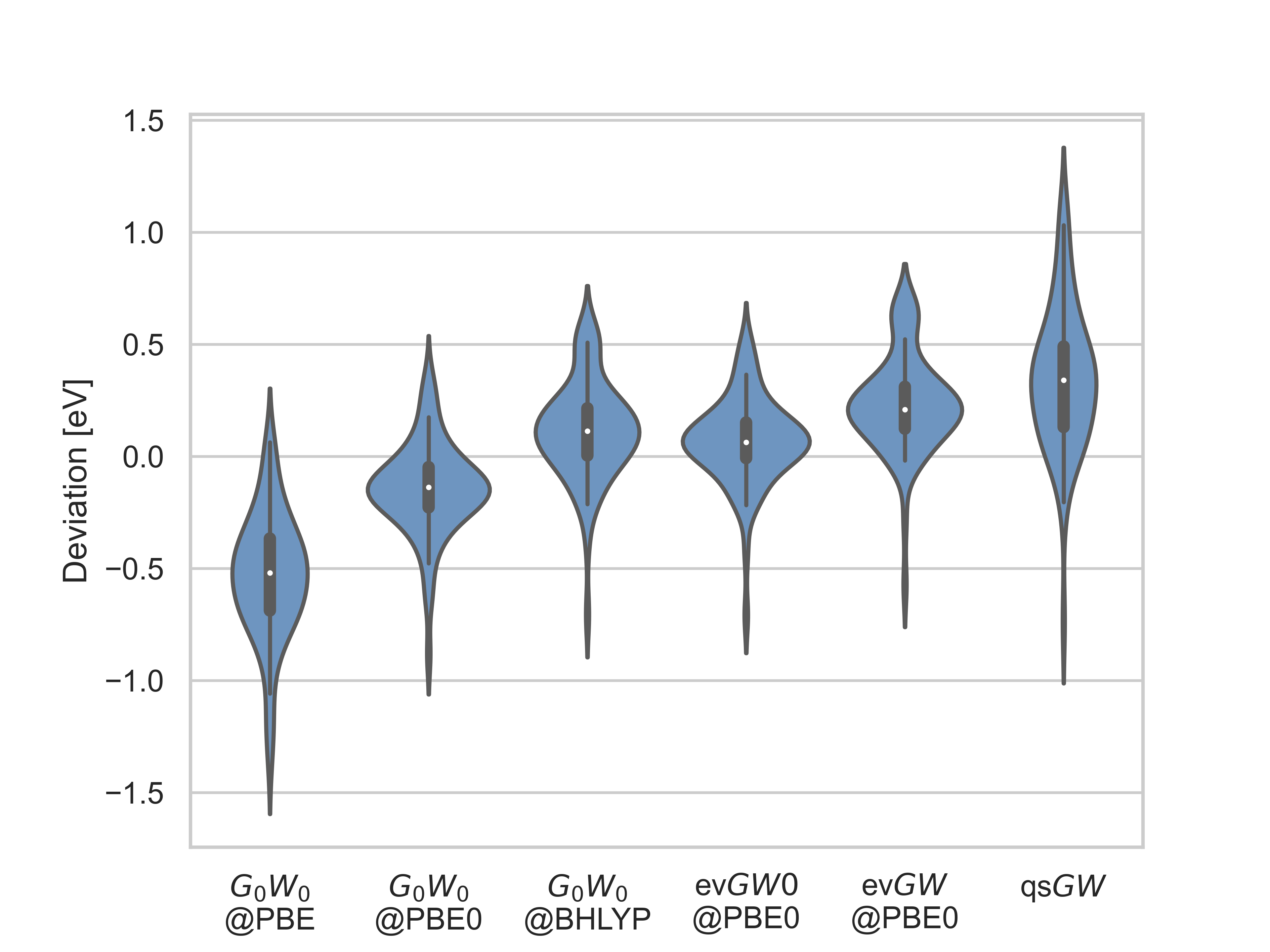}
    \caption{Distribution of the deviations of IPs (in eV) obtained with different 2C methods to the experimental reference values}
    \label{fig::boxes_fr}
\end{figure}

In this section, we compare the different (partially self-consistent) $GW$ variants against experimental IPs. Table~\ref{tab::IPs} shows the first IPs calculated at the 2C level using \eqref{gw_2C} with six different flavors of $GW$: $G_0W_0$ based on PBE, PBE0 and BHLYP orbitals and eigenvalues ($G_0W_0$@PBE, $G_0W_0$@PBE0, $G_0W_0$@BHLYP respectively), ev$GW$ using PBE0 orbitals and eigenvalues (ev$GW$@PBE0), eigenvalue-only self-consistent $GW$ where the screened interaction is fixed at the PBE0 level (ev$GW_0$@PBE0), and qs$GW$. MADs of all considered methods are shown in table~\ref{tab::mads}. The deviations to experiment are also visualized in figure~\ref{fig::boxes_fr}.

Since we take into account SO effects and since our IPs are complete basis set limit extrapolated, vertical experimental IPs are a reliable reference. Besides errors due to the technical parameters discussed in section~\ref{sec::results_west}, other potential sources of uncertainty are the neglect of vibronic effects in our calculations, as well as errors in experimental geometries. Due to the lack of high-quality data from other \emph{ab initio} calculations, these experimental reference values are however the most suitable for our purpose. 

\begin{table}[hbt!]
    \centering
    \begin{tabular}{llcccccc}
    \toprule 
    & & \multicolumn{3}{c}{$G_0W_0$@} & & \\ \cline{3-5}
    & &  PBE & PBE0 & BHLYP &  ev$GW_0$ &  ev$GW$ &  qs$GW$\\ 
    \midrule
    \multirow{3}{*}{MSD} & 
1C-$GW$        & -0.45 & -0.04 & 0.23 & 0.18 & 0.35 & 0.43 \\ 
& 2C-$GW$        & -0.54 & -0.14 & 0.12 & 0.07 & 0.23 & 0.35 \\ 
& 2C-$GW + G3W2$ & -0.46 & -0.06 & 0.22 & 0.15 & 0.35 & 0.47 \\ 
    \midrule
    \multirow{3}{*}{MAD} &
1C-$GW$        & 0.45 & 0.16 & 0.27 & 0.21 & 0.36 & 0.44 \\ 
& 2C-$GW$        & 0.54 & 0.20 & 0.19 & 0.15 & 0.26 & 0.39 \\ 
& 2C-$GW + G3W2$ & 0.46 & 0.14 & 0.25 & 0.20 & 0.37 & 0.49 \\ 
    \bottomrule
    \end{tabular}
    \caption{Mean signed deviations (MSD) and mean absolute deviations (MAD) to experiment for the SOC81* set for different 1C-$GW$, 2C-$GW$ and 2C-$G3W2$ for different starting points and different levels of partial self-consistency. All values are in eV.}
    \label{tab::mads}
\end{table}

Consistent with previous benchmarks on several sets of small and medium molecules,\cite{Marom2012, Bruneval2013, Caruso2016, Knight2016, Zhang2022, Forster2022} $G0W0$@PBE greatly underestimates the first IPs. $G_0W_0$@PBE0 and $G_0W_0$@BHLYP perform much better, with $G_0W_0$@PBE0 showing a tendency to underestimate and $G_0W_0$@BHLYP to overestimate the experimental reference values. BHLYP contains 50 \% of exact exchange which is typically about the optimal fraction for the small and medium organic molecules in the GW100 set.\cite{Zhang2022} The good performance of $G_0W_0$@PBE0 indicates that a smaller fraction of exact exchange might be beneficial for the systems in SOC81*. This might be due to stronger screening effects in these systems containing heavy elements.

In contrast to the cited benchmark studies, ev$GW$ slightly, and qs$GW$ more pronounced, overestimate the reference values. As shown in figure~\ref{fig::boxes_fr}, qs$GW$ is comparable with $G_0W_0$@PBE in showing a larger spread of errors than the best performing methods. The weak performance of this method might be due to the stronger screening in the investigated systems which is typically underestimated by qs$GW$. This then leads to overestimated IPs and HOMO-LUMO gaps. This issue which is well documented for solids\cite{Shishkin2007, VanSchilfgaarde2006,Kang2010,Svane2010,Punya2011} and it has been shown that it can be overcome by inclusion of an effective two-point kernel from time-dependent DFT or the Bethe-Salpeter equation (BSE) with a statically screened exchange kernel.\cite{Tal2021, Cunningham2018,Cunningham2021, Radha2021} Our results indicate that it might be worthwhile to explore such options also for molecular systems. 

With a MAD of 150 meV, the best performing $GW$ method is eigenvalue-only self-consistent $GW$ with the screened interaction kept fixed at the PBE0 level (ev$GW_0$@PBE0). In an ev$GW$ calculation the QP gaps increase during the iterations, leading to underestimated screening. This is compensates for by keeping the screening fixed at the PBE0 level which explains the good performance of this method. It should be noted that despite the partial self-consistency, 2C-ev$GW_0$  is a particularly economic method in our implementation. The 2C polarizability is only to be evaluated once, while the self-energy, which is recalculated in each iteration, is effectively of 1C form. 

\subsubsection{Effect of the perturbative $G3W2$ correction}

\begin{figure}[hbt!]
    \centering
    \includegraphics[width=1.0\textwidth]{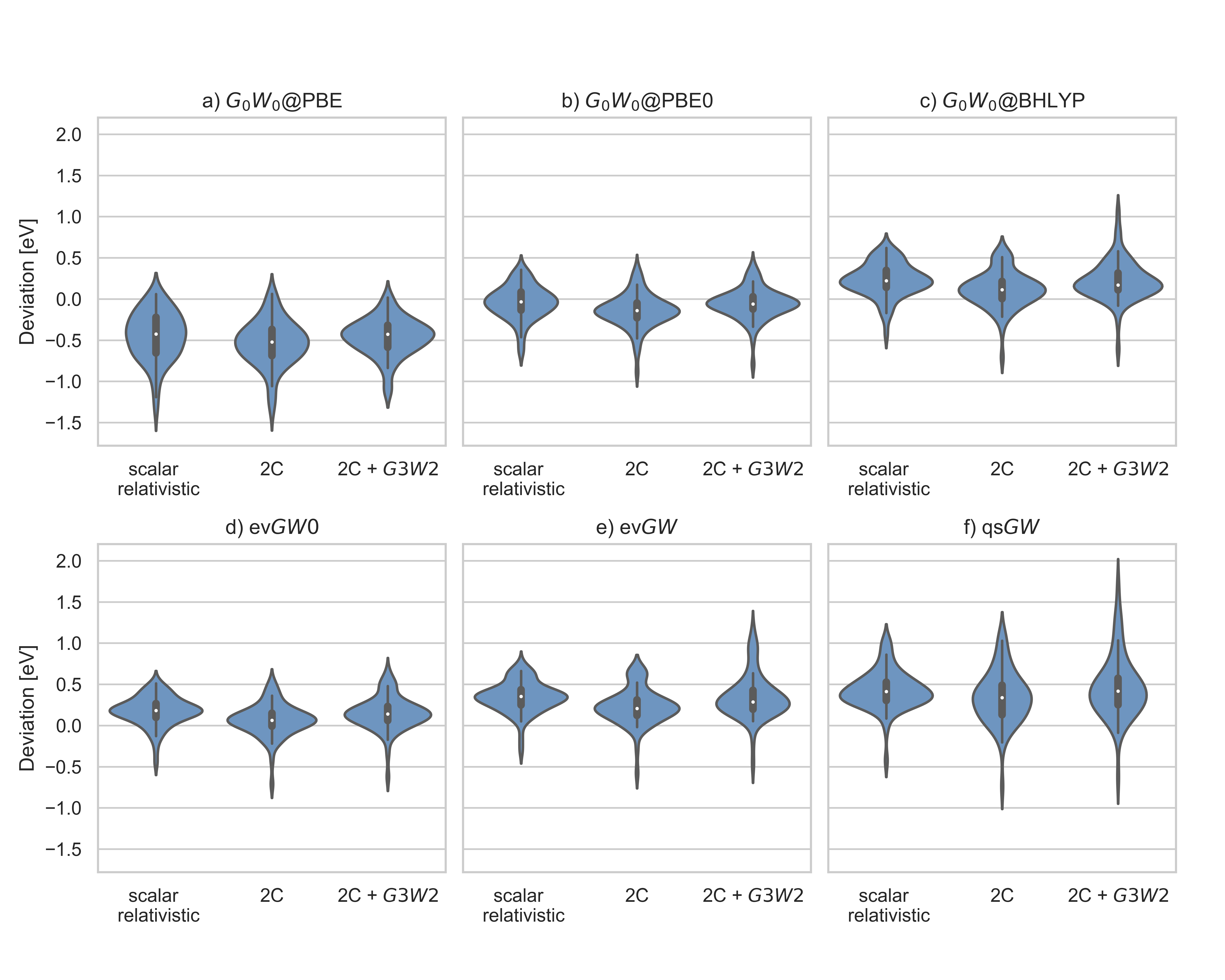}
    \caption{Distribution of the deviations to experimental reference values of IPs. Shown for each method are results for scalar relativistic, 2C and 2C calculations with perturbative $G3W2$ correction. All values are in eV.}
    \label{fig::boxes_both}
\end{figure}

The perturbative inclusion of the $G3W2$ term increases the first IPs. In contrast, in ref.~\citen{Forster2022} it was shown that the $G3W2$ term tends to decrease the IPs in the ACC24 set. As shown in figure~\ref{fig::boxes_both}b), in case of $G_0W_0$@PBE0 the inclusion of this contribution improves agreement with experiment, while for $G_0W_0$@BHLYP and the partially self-consistent methods it worsens it (figure~\ref{fig::boxes_both}c) - f)). Typically, the contribution of the $G3W2$ term to the IP is only of the order of about 0.1 eV. However, in some cases, we observe very large $G3W2$ shifts of up to 0.5 eV, for instance for \ce{RuO4} and \ce{OsO4} for all $GW$ methods. This worsens agreement with experiment but their larger effect underlines the importance of vertex corrections for these systems. Out of all tested methods,  with a MAD of only 140 meV, $G_0W_0$@PBE0~+~$G3W2$ is the most accurate. 

\subsubsection{Shift of ionization potentials due to spin-orbit coupling }

\begin{figure}[hbt!]
    \centering
    \includegraphics[width=0.7\textwidth]{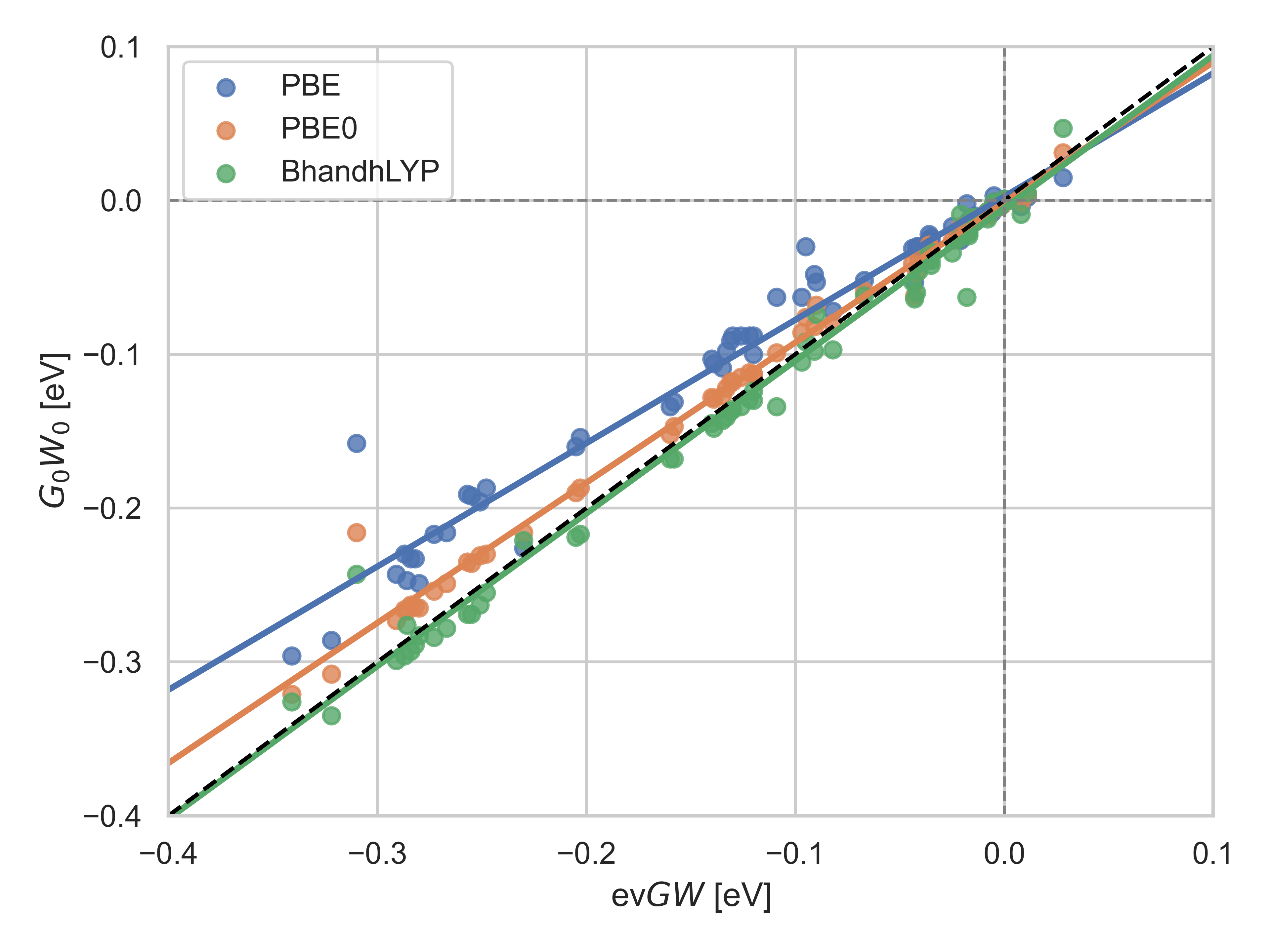}
    \caption{Differences in 2C QP energies to 1C QP energies with $G_0W_0$ using different starting points (x-axis) compared to ev$GW$. All values are in eV.}
    \label{fig::so_splitting}
\end{figure}

Generally, the SOC correction is negative, i.e. reduces the scalar relativistic IPs. This means, in case of $G_0W_0$@PBE0 the scalar relativistic results are in better agreement with experiment than the 2C ones. This is shown infigure~\ref{fig::boxes_both}b). On the other hand, for the accurate partially self-consistent approaches but also for $G0W0$@BHLYP, as shown in figure~\ref{fig::boxes_both}c) to figure~\ref{fig::boxes_both}e), it is crucial to take into account SOC. These observations are also reflected in the MSD and MADs shown in table~\ref{tab::mads}.

Finally, in figure~\ref{fig::so_splitting} we investigate the change in first IPs due to the explicit treatment of SOC among the different $GW$ methods. On the x-axis, we plot the ev$GW$ IPs and on the y-axis the $G_0W_0$ ones for different starting points. A higher amount of exact exchange in the underlying exchange-correlation functional increases the difference between the IPs at the 1C and the 2C level. The same effect as for ev$GW$ can also be observed for qs$GW$ (see supporting information). This can be explained by considering the more (less) pronounced relativistic contraction of the lower (upper) components of a degenerate orbital set that is split by the spin-orbit interaction\cite{Pyykko1979}. The ionization takes place from the upper, more diffuse, orbitals in which the exchange interaction is decreased as compared to the orbitals obtained with a scalar relativistic method. These changes in the exchange interaction induced by relativity are incompletely captured by an approximate exchange density functional approximation resulting in a too small spin-orbit splitting. Employing some non-local exchange, as done in DFT with hybrid functionals, or some form of self-consistency is required to obtain the full magnitude of this subtle effect of relativity. 

\section{\label{sec::conclusions}Conclusions}
We have presented an all-electron, AO based 2C implementation of the GWA for closed-shell molecules in the ADF\cite{adf2022} and BAND\cite{Philipsen2022} engines of AMS\cite{Ruger2022}. As in our 1C $GW$ implementation,\cite{Forster2020b} we leverage the space-time formulation of the GWA, AC of the self-energy, and the PADF approximation to transform between the representations of 4-point correlation functions in the AO and the auxiliary basis to achieve formally cubic scaling with system size.\cite{Forster2020b} The AO-based implementation of the 2C-GWA is particularly efficient: The evaluation of the polarizability is only four times slower than in a 1C calculation. We furthermore only consider the 1-component contribution to the Green's function to evaluate the dynamical part of the self-energy. All in all, this leads to a 2C algorithm which is only about two to three times more expensive than its 1C counterpart. 

While the effect of SOC can faithfully be estimated by combining a 2C DFT calculation with a scalar relativistic $GW$ calculation,\cite{Scherpelz2016} the new implementation will be particularly useful to calculate optical excitations within the 2C-BSE@$GW$ method.

To verify the correctness of our implementation we have calculated the first IPs of a subset of 67 our of the 81 molecules in the SOC81 dataset,\cite{Scherpelz2016} which excludes the multi-solution cases. We have then compared our results to the ones calculated by Scherpelz and Govoni with the WEST code.\cite{Scherpelz2016} For scalar relativistic $G_0W_0$@PBE and $G_0W_0$@PBE0 first IPs, we found MADs to the WEST results of below 100 meV, respectively. With MADs of 70 meV, respectively, the agreement at the 2C level is better than in the scalar relativistic case, which can be rationalized by the different partition of scalar and spin-orbit relativistic effects in both codes. Reaching agreement between $GW$ codes for molecules containing heavy elements is challenging due to relativistic effects and potentially larger errors due to incomplete single particle basis and PPs. As for the GW100 database,\cite{VanSetten2015} further benchmark results using different types of single-particle basis, for instance Gaussian type orbitals, will be necessary to clarify the origin of the discrepancies between both codes. 

Finally, we have used the new implementation to assess the accuracy of $G_0W_0$ based on different starting points and of partially self-consistent approaches for the first IPs of the molecules in the SOC81 set. ev$GW$ and qs$GW$ 
overestimate the experimental vertical ionization energies. Especially the latter method performs poorly, which is in contrast to the good performance for small and medium, predominantly organic molecules\cite{Forster2022, Marie2023}. Both methods are outperformed by $G0W0$ based on PBE0 and BHLYP starting points with fraction of 25 \% and 50\% of exact exchange. With a MAD of 150 meV, out of all $GW$ methods the best agreement with experiment is achieved when the screened interaction is kept fixed at the PBE0 level in an eigenvalue-only self-consistent calculation (ev$GW_0$@PBE0). Including SOC effects though explicit 2C calculations lowers the IPs while the inclusion of the statically screened $G3W2$ correction increases them. Since $G0W0$@PBE0 alone tends to underestimate the experimental reference values, 2C-$G_0W_0$PBE0 + $G3W2$ profits from favorable error cancellation and with a MAD of 140 meV is in excellent agreement with the experimental reference values.

In our benchmarks, we restricted ourselves to 67 out of the 81 molecules in the SOC81 benchmark set. For the other cases the non-linear QP equation \eqref{g0w0-equation} has multiple solutions.\cite{Scherpelz2016} which are difficult to describe correctly with Pade models of the frequency-dependence of the self-energy in an AC treatment. It is important to address this issue, since systems containing heavy elements, including transition metal compounds where problems with AC are ubiquitous, will be among the targets of 2C implementations. AC can be avoided by using analytical integration of the self-energy\cite{Bruneval2012, VanSetten2013, Bintrim2021} or contour deformationtechniques.\cite{Lebegue2003, Govoni2015, Scherpelz2016, Golze2018} AC of the screened interaction can also be combined with CD of the self-energy\cite{Friedrich2019, Duchemin2020} to compute a single-matrix element of the self-energy in the MO basis with cubic scaling with system size. This technique is therefore suitable for $G_0W_0$ and also for ev$GW$ or BSE@$GW$ calculations where Hedin shifts\cite{Pollehn1998, Li2022a} or other rigid scissor-like shifts of the KS spectrum\cite{Vlcek2018b, Holzer2019, Wilhelm2021} can be employed to avoid the explicit calculation of all diagonal elements of the self-energy. Since in qs$GW$ the full self-energy matrix is needed, such an algorithm would scale as $\mathcal{O}\left(N^5\right)$ with system size and is therefore only suitable for small molecules. Together with the already mentioned convergence problems as well as the generally poor performance for the systems considered herein, this is in principle a strong argument against the use of qs$GW$ for such systems.

\appendix 

\section{\label{app::B}Proof of Eqs. 29 and 30}
In this appendix we proof \cref{kramers1,kramers2}, which are valid under Kramers symmetry. We employ relation \cref{kramers-symmetry} to first proof \eqref{kramers2}. In real space,
\begin{equation}
\begin{aligned}
    P^{(0)}(\br \uparrow,\br'\uparrow,i\tau) = & 
    -i \sum_{ia} e^{-\left(\epsilon_a - \epsilon_i\right)\tau}
    \phi_{i}^{\uparrow} (\br)
    \phi_{i}^{\uparrow^*} (\br')
    \phi_{a}^{\uparrow} (\br')
    \phi_{a}^{\uparrow^*} (\br) \\ 
 = & 
    -i \sum_{ia} e^{-\left(\epsilon_a - \epsilon_i\right)\tau}
    \phi_{i}^{\downarrow^*} (\br)
    \phi_{i}^{\downarrow} (\br')
    \phi_{a}^{\downarrow^*} (\br')
    \phi_{a}^{\downarrow} (\br) \\ 
    = & P^{(0)}(\br' \downarrow,\br\downarrow,i\tau) = 
    P^{(0)}(\br \downarrow,\br'\downarrow,i\tau)
\end{aligned}
\end{equation}
with the last equality due to the symmetry of $P^{(0)}$. In the same way, we also show the identity 
\begin{equation}
\begin{aligned}
    P^{(0)}(\br \uparrow,\br'\downarrow,i\tau) = & 
    -i \sum_{ia} e^{-\left(\epsilon_a - \epsilon_i\right)\tau}
    \phi_{i}^{\uparrow} (\br)
    \phi_{i}^{\downarrow^*} (\br')
    \phi_{a}^{\downarrow} (\br')
    \phi_{a}^{\uparrow^*} (\br) \\ 
   = & 
    -i \sum_{ia} e^{-\left(\epsilon_a - \epsilon_i\right)\tau}
    \phi_{i}^{\downarrow^*} (\br)
    \phi_{i}^{\uparrow} (\br')
    \phi_{a}^{\uparrow^*} (\br')
    \phi_{a}^{\downarrow} (\br) \\ 
    = & P^{(0)}(\br' \uparrow,\br\downarrow,i\tau) = 
    P^{(0)}(\br \downarrow,\br'\uparrow,i\tau) \;.
\end{aligned}
\end{equation}
After transformation to the AO basis, these are the identities in \eqref{kramers2}. 

\Cref{kramers1},  
\begin{equation}
\sum_{\sigma, \sigma' = \uparrow,\downarrow}
i G^{>^I}_{\mu \kappa, \sigma  \sigma'}(i\tau)
G^{<^R}_{\nu \lambda, \sigma'  \sigma}(-i\tau) + 
i G^{>^R}_{\mu \kappa,  \sigma  \sigma'}(i\tau)
G^{<^I}_{\nu \lambda, \sigma'  \sigma}(-i\tau) = 0 \;.
\end{equation} 
follows from the cancellation of terms in the sums due to the identities
\begin{align}
\label{appendixB_1}
    G^{>^I}_{\mu\kappa,\uparrow\uparrow}(i\tau)
G^{<^R}_{\nu\lambda, \uparrow\uparrow}(-i\tau) = &
- G^{>^I}_{\mu\kappa, \downarrow \downarrow}(i\tau)
G^{<^R}_{\nu \lambda, \downarrow\downarrow}(-i\tau) \\ 
\label{appendixB_2}
G^{>^R}_{\mu\kappa,\uparrow\uparrow}(i\tau)
G^{<^I}_{\nu\lambda, \uparrow\uparrow}(-i\tau) = &
- G^{>^R}_{\mu\kappa, \downarrow \downarrow}(i\tau)
G^{<^I}_{\nu \lambda, \downarrow\downarrow}(-i\tau) \\ 
\label{appendixB_3}
G^{>^I}_{\mu \kappa, \uparrow \downarrow}(i\tau)
G^{<^R}_{\nu \lambda, \downarrow  \uparrow}(-i\tau) = &
- G^{>^I}_{\mu \kappa, \downarrow \uparrow}(i\tau)
G^{<^R}_{\nu \lambda, \uparrow \downarrow}(-i\tau) \\ 
\label{appendixB_4}
G^{>^R}_{\mu \kappa, \uparrow \downarrow}(i\tau)
G^{<^I}_{\nu \lambda, \downarrow  \uparrow}(-i\tau) = &
- G^{>^R}_{\mu \kappa, \downarrow \uparrow}(i\tau)
G^{<^I}_{\nu \lambda, \uparrow \downarrow}(-i\tau) \;,
\end{align}
These relations follow directly from \cref{greensGWbasisU}, as in each of the four terms there is exactly one sign change upon applying Kramers' symmetry. 

\section{Computational timings}

\sisetup{
  round-mode          = places, 
  round-precision     = 0, 
}
\begin{table}[hbt!]
    \centering
    \begin{tabular}{ll
S[table-format=5.0]%
S[table-format=5.0]%
S[table-format=5.0]%
S[table-format=5.0]%
}
    \toprule 
    & & \multicolumn{2}{c}{{TZ3P}} & \multicolumn{2}{c}{{QZ6P}}  \\ \cline{3-6}
    & & {1C} & {2C} & {1C} & {2C} \\
    \midrule
     {$N_{\text{bas}}$}   & & \multicolumn{2}{c}{1566} & \multicolumn{2}{c}{2895} \\
     {Total}      & {[core h]} &  41  & 82   & 728    &  1995  \\
     {$P^{(0)}$ } & {[core h]}&  14  & 53   & 409    &  1655   \\
     {$W$}        & {[core h]} &   4  &  4   &  30    &    30  \\
     {$\Sigma$}   & {[core h]} &  21  & 20   & 205    &   213  \\
     \midrule
     {first IP}   & {[eV]} & {6.09} & {5.81} &   {6.13} & {5.78} \\ 
     \bottomrule
    \end{tabular}
    \caption{Computational timings and first IP of \ce{Ir(ppy)3} for different basis sets at the 1C and 2C level using $G_0W_0$@PBE0.}
    \label{tab::appendix_timings}
\end{table}

In this appendix we compare the computational timings of 1C and 2C $GW$ calculations in our implementation. We report here timings for Tris(2-phenylpyridine)iridium [\ce{Ir(ppy)3}], a molecule with 320 electrons which is widely used in organic light-emitting diodes (OLEDs) due to its high quantum yields, enabled by thermally activated delayed fluorescence (TADF).\cite{Samanta2017a} Timing results for the full complex at the TZ3P and QZ6P level using the ADF engine are shown in table~\ref{tab::appendix_timings}. Systems like \ce{Ir(ppy)3} which contain many first- and second-row atoms are suitable for AO-based implementations since they can exploit sparsity in the AO basis. For clusters of heavy elements, for instance the \ce{Pb14Se13} cluster considered in ref.~\citen{Scherpelz2016}, MO-based implementations are more suitable, even though their asymptotic scaling with system size is less favorable. 

As one would expect from the equations in section~\ref{sec::theory}, independently of the basis set the calculation of the polarizability is four times slower in the 2C case, while the timings for the other most time-consuming parts of a $G_0W_0$ calculation remain the same. In the QZ calculations, the timings are dominated by the calculation of the polarizability and therefore the 2C calculation is slower compared to the 1C calculation than for the TZ calculations. A single iteration of a partially self-consistent calculation (both ev$GW$ and qs$GW$) is as time-consuming as a $G_0W_0$ calculation. An ev$GW_0$ calculation is more economic as a ev$GW$ calculation, since the polarizability needs to be evaluated only once, saving about a factor of 2 in each iteration.

\begin{acknowledgement}
Edoardo Spadetto acknowledges funding from the European Union's Horizon 2020 research and innovation program under grant agreement No 956813 (2Exciting).
\end{acknowledgement}


\begin{suppinfo}
All Quasiparticle energies calculated in this work. All basis set files.
\end{suppinfo}


\bibliography{all}

\begin{mcitethebibliography}{167}
\bibitem[Not()]{Note-1}
It should be understood that in practice one solves \begin{equation}
  \label{DysonqsGW2} \sum_q\left\{ \left[\Sigma^H_{Hxc}\right]_{pq} -
  \left[\Sigma^{H^{(n-1)}}_{Hxc}\right]_{pq} \right\}\phi_q(\br) =
  \left(\epsilon_p - \epsilon^{(n-1)}_p\right) \phi_p(\br) \end{equation} in
  the $n$th iteration, which reduces to \eqref{DysonqsGW} for $n=1$.\relax
\mciteBstWouldAddEndPunctfalse
\mciteSetBstMidEndSepPunct{\mcitedefaultmidpunct}
{}{\mcitedefaultseppunct}\relax
\EndOfBibitem
\bibitem[Not()]{Note-2}
In principle, there are 15 systems with multiple solutions. However, for
  \ce{CI4}, all three solutions are very close to each other. Therefore, we
  retain this system in our benchmark.\relax
\mciteBstWouldAddEndPunctfalse
\mciteSetBstMidEndSepPunct{\mcitedefaultmidpunct}
{}{\mcitedefaultseppunct}\relax
\EndOfBibitem
\end{mcitethebibliography}


\begin{tocentry}
\includegraphics[width=\textwidth]{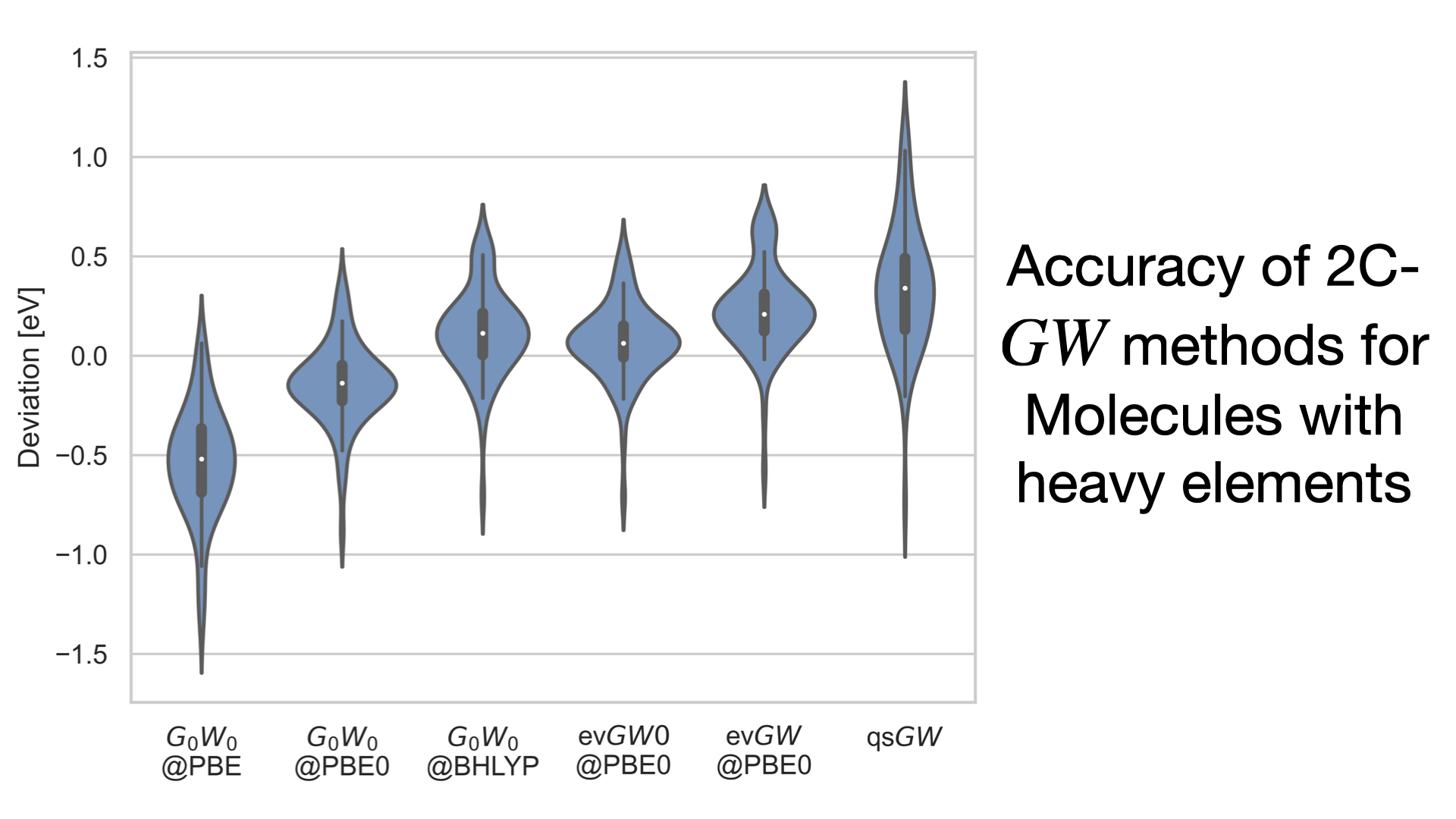}
\end{tocentry}


\end{document}